\documentclass[12pt]{article}

\usepackage{axodraw}

\makeatletter

\def\paragraph{\@startsection{paragraph}{4}{\z@}{+2.00ex plus
 +1ex minus +.2ex}{1.5ex plus .2ex}{\it\normalsize}}

\def\section{\@startsection {section}{1}{\z@}{+3.0ex plus +1ex minus
  +.2ex}{2.3ex plus .2ex}{\normalsize\bf\boldmath}}
\def\subsection{\@startsection{subsection}{2}{\z@}{+2.5ex plus +1ex
minus +.2ex}{1.5ex plus .2ex}{\normalsize\bf\boldmath}}
\def\subsubsection{\@startsection{subsubsection}{3}{\z@}{+3.25ex plus
 +1ex minus +.2ex}{1.5ex plus .2ex}{\normalsize\it}}

 \expandafter\ifx\csname mathrm\endcsname\relax\def\mathrm#1{{\rm #1}}\fi

\newcounter{saveeqn}

\@addtoreset{equation}{section}

\newcount\@tempcntc
\def\@citex[#1]#2{\if@filesw\immediate\write\@auxout{\string\citation{#2}}\fi
  \@tempcnta\z@\@tempcntb\m@ne\def\@citea{}\@cite{\@for\@citeb:=#2\do
    {\@ifundefined
       {b@\@citeb}{\@citeo\@tempcntb\m@ne\@citea
        \def\@citea{,\penalty\@m\ }{\bf ?}\@warning
       {Citation `\@citeb' on page \thepage \space undefined}}%
    {\setbox\z@\hbox{\global\@tempcntc0\csname
b@\@citeb\endcsname\relax}%
     \ifnum\@tempcntc=\z@ \@citeo\@tempcntb\m@ne
       \@citea\def\@citea{,\penalty\@m}
       \hbox{\csname b@\@citeb\endcsname}%
     \else
      \advance\@tempcntb\@ne
      \ifnum\@tempcntb=\@tempcntc
      \else\advance\@tempcntb\m@ne\@citeo
      \@tempcnta\@tempcntc\@tempcntb\@tempcntc\fi\fi}}\@citeo}{#1}}

\def\@citeo{\ifnum\@tempcnta>\@tempcntb\else\@citea
  \def\@citea{,\penalty\@m}%
  \ifnum\@tempcnta=\@tempcntb\the\@tempcnta\else
   {\advance\@tempcnta\@ne\ifnum\@tempcnta=\@tempcntb \else
\def\@citea{--}\fi
    \advance\@tempcnta\m@ne\the\@tempcnta\@citea\the\@tempcntb}\fi\fi}

\def\nl{\nonumber\\}

\newcommand{\gsim}
{\mathrel{\raisebox{-.3em}{$\stackrel{\displaystyle >}{\sim}$}}}
\def\asymp#1%
{\mathrel{\raisebox{-.4em}{$\widetilde{\scriptstyle #1}$}}}

\def\Nequal#1%
{\mathrel{\raisebox{-.5em}{$\stackrel{=}{\scriptstyle\rm#1}$}}}
\newcommand{\dsl}[1]{\not \hspace{-0.7mm}#1}
\def\dsl{\mathpalette\make@slash}
\def\make@slash#1#2{\setbox\z@\hbox{$#1#2$}%
  \hbox to 0pt{\hss$#1/$\hss\kern-\wd0}\box0}

\def\beq{\begin{equation}}
\def\eeq{\end{equation}}
\def\beqar{\begin{eqnarray}}
\def\eeqar{\end{eqnarray}}
\def\barr#1{\begin{array}{#1}}
\def\earr{\end{array}}
\def\bfi{\begin{figure}}
\def\efi{\end{figure}}
\def\btab{\begin{table}}
\def\etab{\end{table}}
\def\bce{\begin{center}}
\def\ece{\end{center}}
\def\nn{\nonumber}
\def\disp{\displaystyle}
\def\text{\textstyle}

\def\arraystretch{1.2}

\def\al{\alpha}
\def\be{\beta}
\def\Ga{\Gamma}
\def\ga{\gamma}
\def\de{\delta}
\def\De{\Delta}
\def\eps{\epsilon}
\def\veps{\varepsilon}
\def\ka{\kappa}
\def\La{\Lambda}
\def\la{\lambda}

\def\si{\sigma}

\def\refeq#1{\mbox{(\ref{#1})}}

\def\reffi#1{\mbox{Figure~\ref{#1}}}
\def\reffis#1{\mbox{Figures~\ref{#1}}}
\def\refta#1{\mbox{Table~\ref{#1}}}
\def\reftas#1{\mbox{Tables~\ref{#1}}}
\def\refse#1{\mbox{Section~\ref{#1}}}
\def\refses#1{\mbox{Sections~\ref{#1}}}

\def\citere#1{\mbox{Ref.~\cite{#1}}}
\def\citeres#1{\mbox{Refs.~\cite{#1}}}

\newcommand{\TeV}{\unskip\,\mathrm{TeV}}
\newcommand{\GeV}{\unskip\,\mathrm{GeV}}

\newcommand{\fb}{\unskip\,\mathrm{fb}}

\newcommand{\ri}{{\mathrm{i}}}
\newcommand{\rd}{{\mathrm{d}}}

\newcommand{\Ord}{\mathswitch{{\cal{O}}}}

\renewcommand{\L}{{\cal{L}}}
\newcommand{\M}{{\cal{M}}}

\def\mathswitchr#1{\relax\ifmmode{\mathrm{#1}}\else$\mathrm{#1}$\fi}

\newcommand{\PV}{V}
\newcommand{\PW}{\mathswitchr W}
\newcommand{\Pw}{\mathswitchr w}
\newcommand{\PZ}{\mathswitchr Z}

\newcommand{\Pg}{\mathswitchr g}
\newcommand{\PH}{\mathswitchr H}
\newcommand{\Pe}{\mathswitchr e}
\newcommand{\Pne}{\mathswitch \nu_{\mathrm{e}}}
\newcommand{\Pnebar}{\mathswitch \bar\nu_{\mathrm{e}}}
\newcommand{\Pnmu}{\mathswitch \nu_{\mu}}
\newcommand{\Pnmubar}{\mathswitch \bar\nu_{\mu}}
\newcommand{\Pd}{\mathswitchr d}
\newcommand{\Pdbar}{\bar{\mathswitchr d}}
\newcommand{\Pu}{\mathswitchr u}
\newcommand{\Pubar}{\bar{\mathswitchr u}}
\newcommand{\Ps}{\mathswitchr s}
\newcommand{\Psbar}{\bar{\mathswitchr s}}
\newcommand{\Pc}{\mathswitchr c}
\newcommand{\Pcbar}{\bar{\mathswitchr c}}

\newcommand{\Pep}{\mathswitchr {e^+}}
\newcommand{\Pem}{\mathswitchr {e^-}}

\newcommand{\Pmup}{\mathswitchr {\mu^+}}
\newcommand{\Pmum}{\mathswitchr {\mu^-}}

\def\mathswitch#1{\relax\ifmmode#1\else$#1$\fi}

\newcommand{\MV}{\mathswitch {M_\PV}}
\newcommand{\MW}{\mathswitch {M_\PW}}

\newcommand{\MZ}{\mathswitch {M_\PZ}}
\newcommand{\MH}{\mathswitch {M_\PH}}

\newcommand{\GW}{\Gamma_{\PW}}
\newcommand{\GZ}{\Gamma_{\PZ}}
\newcommand{\GV}{\Gamma_{\PV}}
\newcommand{\GH}{\Gamma_{\PH}}

\newcommand{\sw}{\mathswitch {s_\Pw}}
\newcommand{\cw}{\mathswitch {c_\Pw}}

\newcommand{\GF}{\mathswitch {G_\mu}}

\def\solid{\raise.9mm\hbox{\protect\rule{1.1cm}{.2mm}}}
\def\dash{\raise.9mm\hbox{\protect\rule{2mm}{.2mm}}\hspace*{1mm}}

\def\ie{i.e.\ }

\def\cf{cf.\ }

\newcommand{\QCD}{{\mathrm{QCD}}}

\newcommand{\SM}{{\mathrm{SM}}}

\newcommand{\CC}{{\mathrm{CC}}}
\newcommand{\NC}{{\mathrm{NC}}}

\newcommand{\had}{\mathrm{had}}
\newcommand{\ew}{\mathrm{ew}}

\newcommand{\az}{a_0}
\newcommand{\ac}{a_{\mathrm{c}}}

\newcommand{\azt}{\tilde a_0}
\newcommand{\aWphi}{\mbox{$\alpha_{\mathrm{W}\phi}$}}
\newcommand{\aBphi}{\mbox{$\alpha_{\mathrm{B}\phi}$}}
\newcommand{\aW}{\mbox{$\alpha_{\mathrm{W}}$}}
\newcommand{\dkag}{\Delta\kappa_{\gamma}}
\newcommand{\lag}{\lambda_{\gamma}}

\newcommand{\U}{\mathrm{U}}
\newcommand{\SU}{\mathrm{SU}}

\def\Re{\mathop{\mathrm{Re}}\nolimits}

\def\lra{\mathop{\mathrm{\leftrightarrow}}\nolimits}

\newcommand{\ggz}{\gamma\gamma\PZ}
\newcommand{\gww}{\gamma\PW\PW}
\newcommand{\gzz}{\gamma\PZ\PZ}
\newcommand{\zzz}{\PZ\PZ\PZ}

\newcommand{\ggww}{\gamma\gamma\PW\PW}
\newcommand{\gzww}{\gamma\PZ\PW\PW}
\newcommand{\ggzz}{\gamma\gamma\PZ\PZ}
\newcommand{\ggvv}{\gamma\gamma\PV\PV}

\newcommand{\ggffff}{\gamma\gamma\to 4f}
\newcommand{\eeffff}{\Pep\Pem\to 4f}
\newcommand{\eeffffff}{\Pep\Pem\to 6f}
\newcommand{\ggffffg}{\ggffff\ga}
\newcommand{\eeffffg}{\eeffff\ga}
\newcommand{\ggffffkg}{\ggffff(\ga)}
\newcommand{\ggtoww}{\gamma\gamma\to \PW\PW}

\newcommand{\ggwwffff}{\gamma\gamma\to \PW\PW\to 4f}

\newcommand{\spab}{\langle k_1 k_2 \rangle}
\newcommand{\spac}{\langle k_1 p_1 \rangle}
\newcommand{\spad}{\langle k_1 p_2 \rangle}
\newcommand{\spae}{\langle k_1 p_3 \rangle}
\newcommand{\spaf}{\langle k_1 p_4 \rangle}
\newcommand{\spbc}{\langle k_2 p_1 \rangle}
\newcommand{\spbd}{\langle k_2 p_2 \rangle}
\newcommand{\spbe}{\langle k_2 p_3 \rangle}
\newcommand{\spbf}{\langle k_2 p_4 \rangle}
\newcommand{\spcd}{\langle p_1 p_2 \rangle}
\newcommand{\spce}{\langle p_1 p_3 \rangle}

\newcommand{\spdf}{\langle p_2 p_4 \rangle}
\newcommand{\spef}{\langle p_3 p_4 \rangle}

\newcommand{\cspab}{\spab^*}
\newcommand{\cspac}{\spac^*}
\newcommand{\cspad}{\spad^*}

\newcommand{\cspaf}{\spaf^*}
\newcommand{\cspbc}{\spbc^*}
\newcommand{\cspbd}{\spbd^*}
\newcommand{\cspbe}{\spbe^*}
\newcommand{\cspbf}{\spbf^*}
\newcommand{\cspcd}{\spcd^*}

\newcommand{\cspdf}{\spdf^*}

\newcommand{\cspef}{\spef^*}

\newcommand{\Ade}{\langle p_2[K_1-P_1]p_3\rangle}
\newcommand{\Afc}{\langle p_4[K_1-P_3]p_1\rangle}

\newcommand{\RacoonWW}{{\sc RacoonWW}}
\newcommand{\Whizard}{{\sc Whizard}}
\newcommand{\Madgraph}{{\sc Madgraph}}
\newcommand{\HDECAY}{{\sc HDECAY}}
\newcommand{\CompAZ}{{\sc CompAZ}}

\def\draftdate{\relax}
\def\mda{\relax}
\def\mua{\relax}
\def\mla{\relax}
\def\draft{
\def\thtystars{******************************}
\def\sixtystars{\thtystars\thtystars}
\typeout{}
\typeout{\sixtystars**}
\typeout{* Draft mode!
         For final version remove \protect\draft\space in source file *}
\typeout{\sixtystars**}
\typeout{}
\def\draftdate{\today}
\def\mua{\marginpar[\boldmath\hfil$\uparrow$]%
                   {\boldmath$\uparrow$\hfil}%
                    \typeout{marginpar: $\uparrow$}\ignorespaces}
\def\mda{\marginpar[\boldmath\hfil$\downarrow$]%
                   {\boldmath$\downarrow$\hfil}%
                    \typeout{marginpar: $\downarrow$}\ignorespaces}
\def\mla{\marginpar[\boldmath\hfil$\rightarrow$]%
                   {\boldmath$\leftarrow $\hfil}%
                    \typeout{marginpar: $\lra$}\ignorespaces}
\def\Mua{\marginpar[\boldmath\hfil$\Uparrow$]%
                   {\boldmath$\Uparrow$\hfil}%
                    \typeout{marginpar: $\uparrow$}\ignorespaces}
\def\Mda{\marginpar[\boldmath\hfil$\Downarrow$]%
                   {\boldmath$\Downarrow$\hfil}%
                    \typeout{marginpar: $\downarrow$}\ignorespaces}
\def\Mla{\marginpar[\boldmath\hfil$\Rightarrow$]%
                   {\boldmath$\Leftarrow $\hfil}%
                    \typeout{marginpar: $\lra$}\ignorespaces}
\overfullrule 5pt
\oddsidemargin -15mm
\marginparwidth 29mm
}

\def\stars{\strut\leaders\hbox{*}\hfill\strut}
\def\starline{\hfil\strut\hfil\hbox to \textwidth {\stars}\hfil}

\begin{document}
\thispagestyle{empty}
\def\thefootnote{\fnsymbol{footnote}}
\setcounter{footnote}{1}
\null
\draftdate\hfill MPP-2004-54\\
\strut\hfill hep-ph/0405169\\
\vspace{1.5cm}
\begin{center}
{\Large \bf\boldmath
Four-fermion production at $\ga\ga$ colliders:
\\[.5em]
1.\ Lowest-order predictions and anomalous couplings
\par} \vskip 2.5em
\vspace{1cm}

{\large
{\sc A.\ Bredenstein, S.\ Dittmaier and M. Roth} } \\[1cm]

{\it Max-Planck-Institut f\"ur Physik (Werner-Heisenberg-Institut)\\
D-80805 M\"unchen, Germany} \\
\par \vskip 1em
\end{center}\par
\vfill
{\bf Abstract:} \par
We have constructed a Monte Carlo generator%
\footnote{The corresponding {\sc Fortran} code can be obtained from the
authors upon request.}
for lowest-order predictions
for the processes $\ggffff$ and $\ggffffg$ in the Standard Model and 
extensions thereof
by an effective $\ga\ga\PH$ coupling as well as
anomalous triple and quartic gauge-boson couplings.
Polarization is fully supported, and a realistic photon beam spectrum
can be taken into account.
For the processes $\ggffff$ all helicity amplitudes are explicitly
given in a compact form.
The presented numerical results contain, in particular, a survey of cross 
sections for representative final states and their comparison to results 
obtained with the program package {\Whizard}/{\Madgraph}. 
The impact of a realistic beam spectrum 
on cross sections and distributions
is illustrated. 
Moreover, the size of various contributions to 
cross sections, such as from weak charged- or neutral-current, or 
from strong interactions, is analyzed. 
Particular attention is paid to W-pair production channels
$\ga\ga\to\PW\PW\to 4f(\ga)$ where we investigate the impact of background
diagrams, possible definitions of the W-pair signal, and 
the issue of gauge-invariance violation caused by 
finite gauge-boson widths.
Finally, the effects of triple and quartic anomalous gauge-boson couplings
on cross sections as well as the possibility to constrain these
anomalous couplings at future $\ga\ga$ colliders are discussed.
\par
\vskip 1cm
\noindent
May 2004
\null
\setcounter{page}{0}
\clearpage
\def\thefootnote{\arabic{footnote}}
\setcounter{footnote}{0}

\section{Introduction}
\label{se:intro}

A photon (or $\ga\ga$) collider \cite{Ginzburg:1981vm} as an option at 
a future $\Pep\Pem$ linear collider extends the physics potential of such a 
machine substantially. 
It provides us with information about new physics phenomena, 
such as properties of the Higgs boson or of new particles, 
which is in many respects complementary in the 
$\Pep\Pem$ and $\ga\ga$ modes 
(see, e.g., \citeres{Ginzburg:1981vm,DeRoeck:2003gv} and references therein). 
For instance, Higgs-boson production in the $s$-channel 
\cite{Ginzburg:1981vm,DeRoeck:2003gv,Belanger:1992qi,Boos:1991id}
becomes accessible which is highly suppressed in $\Pep\Pem$ annihilation, but
very interesting for Higgs bosons of relatively large mass.
Owing to its large cross section, W-pair production
is of particular interest at a $\ga\ga$ collider. 
For instance, it can be used for precision 
tests of the gauge sector of the Electroweak Standard Model (SM). 
While the reaction $\Pep\Pem\to\PW\PW$ 
depends on the gauge-boson couplings $\PZ\PW\PW$ and $\gww$, 
the corresponding reaction at a $\ga\ga$ collider, $\ggtoww$, is sensitive 
to the gauge-boson couplings $\gww$ and $\ggww$. 
At an $\Pep\Pem$ collider the coupling $\ggww$ is only directly 
accessible through the bremsstrahlung process $\Pep\Pem\to\PW\PW\ga$ 
which is suppressed by a factor $\al(0)/\pi$
w.r.t.\ the non-radiative process $\Pep\Pem\to\PW\PW$.
Therefore, the sensitivity to the anomalous $\ggww$ coupling in the 
$\ga\ga$ mode is expected to be an order of magnitude better 
than in the $\Pep\Pem$ mode. The precision for the measurement 
of the $\gww$ coupling is comparable in both modes 
\cite{Bozovic-Jelisavcic:2002ta}.

Since W~bosons decay into fermion--anti-fermion pairs, the actually
observed final states of $\ggtoww$ involve four fermions. 
Lowest-order predictions for $\ggffff$ processes (with monochromatic
photon beams and leptonic or semileptonic final states) were
discussed in \citeres{Moretti:1996nv,Baillargeon:1997se}.
With the exception of \citere{Baillargeon:1997rz},
the existing analyses on non-standard couplings at a $\ga\ga$ collider, 
which focus on anomalous triple gauge-boson couplings (ATGC) 
\cite{Bozovic-Jelisavcic:2002ta,Tupper:1980bw},
on anomalous quartic gauge-boson couplings (AQGC) 
\cite{Belanger:1992qi,Marfin:2003jg},
on CP-violating gauge-boson couplings \cite{Belanger:hp},
and on effects of strongly interacting longitudinal W~bosons
\cite{Herrero:1992zq},
treat W~bosons as stable.
In the above studies
radiative corrections were not properly taken into account either.

In view of the achievable experimental precision, 
the theoretical predictions should include, however, both radiative 
corrections and
the decays of the W~bosons, i.e., the full process $\ggffff$ should 
be considered.
The lowest-order amplitudes to the processes $\ggffff$ include 
diagrams with two resonant W bosons (``signal diagrams'') 
as well as diagrams with only one or no W~resonance (``background diagrams'').
Compared to the doubly-resonant diagrams, such singly-resonant and 
non-resonant diagrams are suppressed by 
roughly a factor $\GW/\MW$ and $(\GW/\MW)^2$, respectively.
On the other hand, radiative corrections modify the theoretical predictions
by contributions of ${\cal O}(\alpha)$.
Since $\GW/\MW$ 
is also of ${\cal O}(\alpha)$, 
both the off-shell effects and the
radiative corrections 
contribute to the $\ga\ga\to\PW\PW\to 4f$ cross section in the same 
perturbative order.
Hence, it is a promising approach to calculate the lowest-order matrix 
elements to the full $\ggffff$ process and take into account 
only the doubly-resonant part of the ${\cal O}(\alpha)$ corrections to 
the reaction $\ga\ga\to\PW\PW\to 4f$.
This approach is known as double-pole approximation (DPA) where only 
the leading terms in the expansion of the amplitude around the 
poles of the resonant W propagators are included in the calculation. 
The overall accuracy is then ${\cal O}(\alpha)$ relative to the
lowest-order prediction for
the doubly-resonant W-pair signal.
This approach has been described in \citere{Aeppli:1993rs} in some detail
and successfully applied to W-pair production in $\Pep\Pem$ annihilation
in different versions
\cite{Jadach:1998tz,Beenakker:1998gr,Denner:1999gp,Denner:2000kn,Kurihara:2001um}.

The doubly-resonant virtual corrections consist of
factorizable and non-factorizable contributions. 
The former comprise the corrections to on-shell W-pair 
production \cite{Denner:1994ca,Denner:1995jv,Jikia:1996uu}%
\footnote{Radiative corrections to on-shell W-pair production,
$\ga\ga\to\PW\PW$, were also considered in \citere{Marfin:2003yq}.}
and the decay of on-shell W~bosons \cite{Bardin:1986fi},
the latter account for soft-photon exchange between the production
and decay subprocesses \cite{Melnikov:1995fx}.
Although the basic building blocks for the virtual corrections exist 
in the literature, the combination into 
a complete set of 
${\cal O}(\alpha)$ corrections in DPA has not been done yet.
The evaluation of bremsstrahlung 
processes $\ggffff+\ga$ considered in this work 
serves as a further building block for the ${\cal O}(\alpha)$
corrections in DPA. 

The calculation of lowest-order predictions for the processes $\ggffff$ 
and $\ggffffg$ described in this paper closely follows the approach 
of \citere{Denner:1999gp} for the $\Pep\Pem$ case.
Fermion masses are neglected everywhere assuming that 
all mass singularities are avoided by phase-space cuts.
Following \citere{Denner:1999gp} we construct an event 
generator based on multi-channel Monte Carlo integration 
\cite{Berends:1994pv} using adaptive self-optimization \cite{Kleiss:qy}.
A realistic photon spectrum
can optionally be included in the 
parametrization of {\CompAZ} \cite{Zarnecki:2002qr} which is based 
on the results of \citere{Telnov:hc}.
In addition to the pure SM case, we calculate also the lowest-order matrix 
elements including anomalous couplings. Specifically, we consider 
ATGC \cite{Gaemers:1978hg,Gounaris:1996rz} 
corresponding to SU(2)$\times$U(1)-gauge-invariant 
dimension-6 operators and genuine AQGC 
\cite{Belanger:1992qh,Denner:2001vr} which respect electromagnetic 
gauge invariance and custodial $\SU(2)_{\mathrm{c}}$ symmetry.
Finally, we include an effective $\ga\ga\PH$ coupling in our analysis
in order to study the Higgs resonance in 
the reaction $\ga\ga\to\PH\to\PW\PW/\PZ\PZ\to 4f$.

The performance of the Monte Carlo generator is illustrated in a detailed 
numerical discussion. We present a survey of cross sections for 
a set of representative $4f$ and $4f\ga$ final states and compare them
with results obtained by the combination of the programs
{\Whizard} \cite{Kilian:2001qz} version 1.28
and {\Madgraph} \cite{Stelzer:1994ta}.
We find agreement within statistical uncertainties.
Moreover, we discuss some invariant-mass,
energy, and angular distributions.
We illustrate the impact of the realistic beam spectrum
on cross sections and distributions, and study the size of various 
contributions to cross sections, such as from weak charged- or 
neutral-current, or from strong interactions. 
In view of the aimed precision calculation for 
$\ga\ga\to\PW\PW\to 4f(\ga)$, which will include radiative corrections
in DPA, we elaborate on the possibility
to define a proper W-pair production signal. In this context,
we investigate a naive signal definition which is based on doubly-resonant 
diagrams only and thus violates gauge invariance. We find that this naive
definition, which works well for the $\Pep\Pem$ case
(cf.\ discussion of the so-called 
``CC03'' cross section
in \citere{Grunewald:2000ju}), is not satisfactory for
$\ga\ga\to\PW\PW\to 4f$, but requires a proper gauge-invariant definition
of the residue on the double resonance.
For some $\ggffff$ and 
$\ggffff\ga$ processes we investigate
the effects of gauge-invariance violation by introducing gauge-boson
decay widths. 
To this end, we compare results obtained 
with the naive introduction
of constant or running widths with the gauge-invariant result obtained
with the so-called ``complex-mass scheme'' \cite{Denner:1999gp}.
Finally, we discuss the effects of ATGC and AQGC on cross sections.
In particular, we estimate the bounds on 
anomalous couplings
that could be set by a $\ga\ga$ collider, by assuming 
a photon spectrum
and luminosities expected for the $\ga\ga$ option at TESLA.

The paper is organized as follows. In \refse{se:helamp} we 
describe the calculation of the helicity amplitudes in the SM and 
give explicit results for $\ggffff$. In \refses{se:agc} and \ref{se:higgs} 
we describe the calculation of 
the amplitudes with anomalous gauge-boson couplings 
and an effective $\ga\ga\PH$ interaction. 
\refse{sec:integration} provides some details about the numerical Monte 
Carlo integration. In \refse{se:numerics} we present the numerical results 
as outlined above. Our summary 
is given in \refse{se:sum}.

\section{Analytical results for amplitudes in the Standard Model}
\label{se:helamp}
\subsection{Notation and conventions}
\label{se:not&con}

We consider reactions of the types
\beqar\label{eq:ggffff}
\ga(k_1,\la_1)+\ga(k_2,\la_2) &\to& 
f_1(p_1,\si_1)+\bar f_2(p_2,\si_2)+f_3(p_3,\si_3)+\bar f_4(p_4,\si_4),
\qquad\\*
\gamma(k_1,\la_1)+\gamma(k_2,\la_2) &\to& 
f_1(p_1,\si_1)+\bar f_2(p_2,\si_2)+f_3(p_3,\si_3)+\bar f_4(p_4,\si_4)
+\gamma(p_5,\si_5).\qquad
\label{eq:ggffffg}
\eeqar
The arguments label the momenta $k_i$, $p_j$ and helicities
$\lambda_k$, $\si_l$ (which take the values $\pm1/2$ in the case of fermions and 
$\pm1$ in the case of photons) of the corresponding particles. We often
use only the signs to denote the helicities. The fermion masses are
neglected everywhere.

For the Feynman rules we follow the conventions of \citere{Bohm:rj}.
We extend the usual linear gauge-fixing term in the 
't~Hooft--Feynman gauge by a non-linear part according to 
\citeres{Denner:1999gp,Denner:1995jv,Fujikawa:qs}
such that the vertex 
$\ga\PW\phi$ vanishes, where 
$\phi$ are the would-be Goldstone bosons corresponding to
the W~bosons. 
Note that this also affects the gauge-boson couplings 
$\ga\ga\PW\PW$ and $\ga\PW\PW$.
The corresponding Feynman rules relevant for $\ggffff(\ga)$ in lowest
order can be found in \citere{Denner:1999gp}.
Since we neglect fermion masses,
the would-be Goldstone bosons 
do not couple to fermions and do not occur in the Feynman graphs
of the SM amplitudes to $\ggffff(\ga)$, which leads to a considerable 
reduction of the number of Feynman diagrams. 
 
\subsection{Classification of final states for $\ggffff(\ga)$}
\label{se:pclass}

The final states for $\ggffff$ and $\ggffffg$ can be classified similarly 
to the processes $\eeffff$ and $\eeffffg$ \cite{Denner:1999gp}.
In the following, $f$ and $F$ are 
different fermions ($f\ne F$), and $f'$ and $F'$ denote their 
weak-isospin partners, respectively.
We distinguish between states that are produced via charged-current (CC), 
via neutral-current (NC) interactions, or via both interaction types:

\renewcommand{\labelenumi}{(\roman{enumi})}
\renewcommand{\labelenumii}{(\alph{enumii})}
\newcommand{\ientry}[2]{\mbox{\rlap{#1}\hspace*{5cm}#2}}
\begin{enumerate}
\item CC reactions:
\begin{enumerate}
\item[] \ientry{$\gamma\gamma \to f \bar f' F \bar F'$}%
             ({\em CC31} family), 
\end{enumerate}
\item NC reactions:
\begin{enumerate}
\item \ientry{$\gamma\gamma \to f \bar f F \bar F$}%
             {({\em NC40} family),}
\item \ientry{$\gamma\gamma \to f \bar f f \bar f$}%
             {({\em NC2$\cdot$40} family),}
\end{enumerate}
\item Mixed CC/NC reactions:
\begin{enumerate}
\item[] \ientry{$\gamma\gamma \to f \bar f f' \bar f'$}%
             {({\em mix71} family).}
\end{enumerate}
\end{enumerate}
The radiation of an additional photon does not change this
classification.
Following \citere{CERN9601mcgen} we give the names of the process 
families in parentheses where the numbers correspond to the 
number of Feynman diagrams involved 
in unitary or non-linear gauge (for processes without neutrinos 
in the final state, not counting gluon-exchange diagrams).

Since the matrix elements depend on the colour structure of 
the final state we further distinguish between leptonic, 
semi-leptonic, and hadronic final states. Keeping in mind that we 
neglect fermion masses,
omitting four-neutrino final states, and suppressing reactions that are 
equivalent by CP symmetry we end up with 17 different representative 
processes which we have listed in \refta{ta:classes}.

\begin{table}[ht]
\renewcommand{\arraystretch}{1.1}
\centerline{
\begin{tabular}{|c|c|c|}
\hline
final state & reaction type & $\ga\ga\to$ \\
\hline
leptonic & CC & $\Pem\Pnebar\Pnmu\Pmup$ \\
\cline{2-3}
& NC(a) & $\Pem\Pep\Pnmu\Pnmubar$ \\
\cline{3-3}
& & $\Pem\Pep\Pmum\Pmup$ \\
\cline{2-3}
& NC(b) & $\Pem\Pep\Pem\Pep$ \\
\cline{2-3}
& CC/NC & $\Pem\Pep\Pne\Pnebar$ \\
\cline{1-3}
semi-leptonic & CC(c) & $\Pem\Pnebar\Pu\Pdbar$ \\
\cline{2-3}
& NC(a) & $\Pne\Pnebar\Pu\Pubar$ \\
\cline{3-3} 
&& $\Pne\Pnebar\Pd\Pdbar$ \\
\cline{3-3} 
&& $\Pem\Pep\Pu\Pubar$ \\
\cline{3-3} 
&& $\Pem\Pep\Pd\Pdbar$ \\
\cline{1-3} 
hadronic & CC & $\Pu\Pdbar\Ps\Pcbar$ \\
\cline{2-3}
& NC(a) & $\Pu\Pubar\Pc\Pcbar$ \\
\cline{3-3}
& NC(a) & $\Pu\Pubar\Ps\Psbar$ \\
\cline{3-3}
& NC(a) & $\Pd\Pdbar\Ps\Psbar$ \\
\cline{2-3}
& NC(b) & $\Pu\Pubar\Pu\Pubar$ \\
\cline{3-3}
& NC(b) & $\Pd\Pdbar\Pd\Pdbar$ \\
\cline{2-3}
& CC/NC & $\Pu\Pubar\Pd\Pdbar$ \\
\hline
\end{tabular}}
\caption{Set of representative processes for $\ggffffkg$.}
\label{ta:classes}
\end{table}
Since the photons are polarized after 
Compton backscattering, 
final states that are flavour equivalent up to a CP transformation 
need not necessarily yield the same cross section if the convolution
over a realistic photon beam spectrum 
is included. However, as we neglect 
fermion masses, this is only relevant for the 
semileptonic CC processes 
$\ga\ga\to\Pem\Pnebar\Pu\Pdbar(\ga)$ and $\ga\ga\to\Pne\Pep\Pd\Pubar(\ga)$. 

\subsection{Lowest-order amplitudes for \boldmath{$\ga\ga\to 4f$} }
\label{se:ee4f}
 
\subsubsection{Construction of matrix elements}

The amplitudes for the processes $\ggffff$ are constructed by 
attaching the two incoming photons in all possible ways to the 
corresponding diagrams with 
four external fermions as shown in \reffi{fig:gengraph}.
\bfi
\centerline{
\begin{picture}(120,130)(0,0)
\GOval(75,55)(43,22)(0){0.5}
\ArrowLine(75,30)(105,45)
\ArrowLine(105,15)(75,30)
\Vertex(75,30){2}
\put(108,43){$f_3$}
\put(108,8){$\bar f_4$}
\Photon(75,30)(75,80){2}{6}
\put(62,53){$\PV$}
\Vertex(75,80){2}
\ArrowLine(75,80)(105,95)
\ArrowLine(105,65)(75,80)
\put(108,93){$f_1$}
\put(108,63){$\bar f_2$}
\Photon(24,40)(54,45){2}{4}
\Photon(24,70)(54,65){2}{4}
\put(7,35){$\ga_2$}
\put(7,70){$\ga_1$}
\end{picture}}
\caption{Generic diagram for the process $\ggffff$ where the photons 
$\ga_1,\ga_2$ couple to the fermions $f_1,\dots,\bar f_4$ and the 
gauge boson $\PV$ in all possible ways.}
\label{fig:gengraph}
\efi
The matrix element of the generic diagram in \reffi{fig:gengraph},
where two fermion lines are linked by a gauge boson $V$,
can be written as
\beqar\label{genfun4f}
\M_{\la_1\la_2,\PV}^{\si_1\si_2\si_3\si_4}
(k_i,p_j,Q_j) 
&=& 4e^4 \de_{\si_1,-\si_2} \de_{\si_3,-\si_4} \, 
g^{\si_1}_{\PV\bar f_1 f_2} g^{\si_3}_{\PV\bar f_3 f_4}
A^{\si_1\si_3}_{\la_1\la_2,\PV}(k_i,p_j,Q_j),
\eeqar
where $k_i$, $p_j$ and $Q_j$ ($i=1,2;j=1,..,4$) stand for the momenta 
and relative electric charges of the particles, respectively. 
The coupling factors 
(relative to the electric unit charge $e$) read 
\beqar
\begin{array}[b]{lcllcl}
g^\si_{\ga\bar f_i f_i} &=& -Q_i,&
g^\si_{\PZ\bar f_i f_i} &=& \disp
-\frac{\sw}{\cw}Q_i+\frac{I^3_{{\mathrm{w}},i}}{\cw\sw}\delta_{\si-}, 
\\
g^\si_{\PW\bar f_i f'_i} &=& \disp\frac{1}{\sqrt{2}\sw}\delta_{\si-},\qquad&
g^{\si}_{\Pg\bar f_i f_i} &=& \disp\frac{g_{\mathrm{s}}}{e},
\end{array}
\eeqar
where $I^3_{{\mathrm{w}},i}=\pm{1/2}$ denotes 
the weak isospin of the fermion $f_i$ and $g_{\mathrm{s}}$
the strong coupling constant. The weak mixing angle is defined by 
\beq
\cos{\theta_\mathrm{w}}=\cw=\frac{\MW}{\MZ},\qquad \sw=\sqrt{1-\cw^2}.
\eeq
Quark mixing is neglected everywhere, i.e.\ we set the CKM matrix
equal to the unit matrix.
The auxiliary functions $A^{\si_1\si_3}_{\la_1\la_2,\PV}$
are calculated within the
Weyl--van-der-Waerden (WvdW) formalism following the conventions of 
\citere{Dittmaier:1998nn}.
The WvdW spinor products are defined by
\beq
\langle pq\rangle=\epsilon^{AB}p_A q_B
=2\sqrt{p_0 q_0} \,\Biggl(
{\mathrm{e}}^{-\ri\phi_p}\cos\frac{\theta_p}{2}\sin\frac{\theta_q}{2}
-{\mathrm{e}}^{-\ri\phi_q}\cos\frac{\theta_q}{2}\sin\frac{\theta_p}{2}
\Biggr), 
\eeq
where $p_A$, $q_A$ are the associated momentum spinors for the momenta
\beqar
p^\mu&=&p_0(1,\sin\theta_p\cos\phi_p,\sin\theta_p\sin\phi_p,\cos\theta_p),\nl
q^\mu&=&q_0(1,\sin\theta_q\cos\phi_q,\sin\theta_q\sin\phi_q,\cos\theta_q).
\eeqar 
Moreover, we define the shorthands
\beqar
\langle p_iP_kp_j\rangle &=& p_{i,\dot A}P_k^{\dot AB}p_{j,B} =
p_{i,\dot A}p_k^{\dot A}p_k^Bp_{j,B}=\langle p_ip_k\rangle^*
\langle p_jp_k\rangle,
\nn\\[.3em]
\langle p_i[P_l+P_m]p_j\rangle &=& \langle p_iP_lp_j\rangle 
+\langle p_iP_mp_j\rangle,
\eeqar
where $p_{k,l,m}$ are light-like momenta, i.e., $p_k^2=p_l^2=p_m^2=0$.
In the following,
the denominators of the gauge-boson propagators are abbreviated by
\beq
P_{\PV}(p) = \frac{1}{p^2-\MV^2}, \qquad \PV=\ga,\PZ,\PW,\Pg, 
\qquad M_{\ga} = M_{\Pg} = 0.
\eeq
The introduction of the finite width is described in 
\refse{se:finwidth} below.

The auxiliary functions $A^{\si_1\si_3}_{\la_1\la_2,\PV}$
explicitly read
\beqar
\lefteqn{A^{--}_{++,\PV}(k_i,p_j,Q_j)=(\cspdf)^2} &&
\nn\\
&& {}
\times\Biggl\{-Q_1^2\frac{\cspcd\spef P_{\PV}(p_3+p_4)}
{\cspac\cspad\cspbc\cspbd}
-Q_1Q_3\frac{(p_1+p_2-k_1)^2P_{\PV}(p_1+p_2-k_1)}{\cspac\cspad\cspbe\cspbf}
\nn\\
&& \qquad {}
+Q_3(Q_1-Q_2)P_{\PV}(p_1+p_2)
\nn\\
&& \qquad\qquad {}
\times\Biggl[ \frac{-\cspdf\spcd+\cspaf\spac\MV^2
P_{\PV}(p_1+p_2-k_1)}{\cspad\cspaf\cspbe\cspbf}
+(k_1\lra k_2)\Biggr]
\nn\\
&& \qquad {}+(Q_1-Q_2)^2P_{\PV}(p_1+p_2)P_{\PV}(p_3+p_4)
\left[-\cspdf\frac{\cspdf\spcd\spef+\MV^2\spce}
{2\cspad\cspaf\cspbd\cspbf}\right.
\nn\\
&& \qquad\qquad {}
+\left.\MV^2P_{\PV}(p_1+p_2-k_1)
\frac{\spac\spbe}{\cspad\cspbf}\right]
\nn\\
&& {}
+\Big(\{p_1,Q_1;p_2,Q_2\} \lra \{p_3,Q_3;p_4,Q_4\}\Big)\Biggr\} ,
\nn\\[1em]
\lefteqn{A^{--}_{+-,\PV}(k_i,p_j,Q_j)=
Q_1^2P_{\PV}(p_3+p_4)\frac{\cspdf\spac\langle k_2[P_2+P_4]p_3\rangle}{\cspac\spbc
(p_2+p_3+p_4)^2}} &&
\nn\\
&& {}
+Q_2^2P_{\PV}(p_3+p_4)\frac{\cspbd\spce\langle p_4[P_1+P_3]k_1\rangle}
{\cspad\spbd(p_1+p_3+p_4)^2} 
\nn\\
&& {}
+Q_1Q_2P_{\PV}(p_3+p_4)\frac{\Ade\Afc}
{\cspac\cspad\spbc\spbd} 
\nn\\
&& {}
+(Q_2-Q_1)P_{\PV}(p_3+p_4) \frac{\cspdf\spce}{\cspad\spbc}
\left[ Q_2\frac{\Afc}{\cspaf\spbd} +Q_1\frac{\Ade}{\cspac\spbe} \right]
\nn\\
&& {}
+\frac{1}{2}(Q_2-Q_1)^2P_{\PV}(p_1+p_2)P_{\PV}(p_3+p_4)
\nn\\
&& \qquad {} \times\frac{\cspdf\spce
\Big(\Ade\Afc-M_{\PV}^2\cspdf\spce\Big)}{\cspad\cspaf\spbc\spbe}
\nn\\
&& {}
+\left[-Q_1+(Q_1-Q_2)2(k_1p_1)P_{\PV}(p_1+p_2)\right]
\left[Q_4+(Q_3-Q_4)2(k_2p_4)P_{\PV}(p_3+p_4)\right]
\nn\\
&& \qquad {} \times\frac{(\Ade)^2P_{\PV}(p_1+p_2-k_1)}{\cspac\cspad\spbe\spbf}
\nn\\
&& {}
+\Big(\{p_1,Q_1;p_2,Q_2\} \lra \{p_3,Q_3;p_4,Q_4\}\Big).
\label{eq:24ampl}
\eeqar
The other auxiliary functions $A^{\si_1\si_3}_{\la_1\la_2,\PV}$ 
follow from the relations
\beqar
\label{eq:CP}
A^{-\si_1,\si_3}_{\phantom{-}\la_1\la_2,\PV}(k_i,p_j,Q_j)=
\left[A^{\si_1\si_3}_{\la_1\la_2,\PV}(k_i,p_j,Q_j)
\right]_{\{p_1,Q_1\} \lra \{p_2,-Q_2\}},\nn \\
A^{\si_1,-\si_3}_{\la_1\la_2,\PV}(k_i,p_j,Q_j)=
\left[A^{\si_1\si_3}_{\la_1\la_2,\PV}(k_i,p_j,Q_j)
\right]_{\{p_3,Q_3\} \lra \{p_4,-Q_4\}}
\eeqar
and
\beq
\label{eq:P}
A^{-\si_1,-\si_3}_{-\la_1,-\la_2,\PV}(k_i,p_j,Q_j)=
\left[A^{\si_1\si_3}_{\la_1\la_2,\PV}(k_i,p_j,Q_j)\right]^*.
\eeq
The last relation expresses a parity transformation.
Note that the operation of complex conjugation in Eq.~\refeq{eq:P} 
must not affect the gauge-boson widths in the propagator
functions $P_V$ which will be introduced in \refse{se:finwidth}.

The calculation of the helicity amplitudes for $\ggffff\ga$ proceeds
along the same lines. The result, however, is quite lengthy so that we
do not write it down explicitly.

\subsubsection{Squared amplitudes for leptonic and semi-leptonic final states}
\label{se:lepfinstat}

The result for leptonic and semi-leptonic final states follows immediately 
from the generic amplitude \refeq{genfun4f}. 
The gauge boson cannot be a gluon in this case, and
the sum over the colour degrees of freedom 
in the squared matrix elements trivially leads to the global factors 
$N^{\mathrm{c}}_{\mathrm{lept}}=1$ and 
$N^{\mathrm{c}}_{\mathrm{semilept}}=3$.
Note that for NC diagrams the result for the amplitude
is much simpler than for CC diagrams,
since all terms in Eq.~\refeq{eq:24ampl} 
involving a factor ($Q_1-Q_2$) or ($Q_3-Q_4$) drop out. 
Most of these terms
originate from diagrams where a photon couples 
to a virtual W~boson.

The explicit results for the 
colour-summed squared matrix elements read
\beqar
\sum_{\mathrm{colour}}|{\cal M}_{\CC}|^2 &=&
N^{\mathrm{c}}|{\cal M}_{\PW}|^2,
\label{eq:MCC2}
\\
\sum_{\mathrm{colour}}|{\cal M}_{\NC(\mathrm{a})}|^2 &=&
N^{\mathrm{c}}\left|\M_\NC\right|^2,
\\ 
\sum_{\mathrm{colour}}|{\cal M}_{\NC(\mathrm{b})}|^2 &=&
N^{\mathrm{c}}\left|\M_\NC
-\left[\M_\NC
\right]_{\{p_1,Q_1,\si_1\} \lra \{p_3,Q_3,\si_3\}}\right|^2,
\\ 
\sum_{\mathrm{colour}}|{\cal M}_{\CC/\NC}|^2 &=&
N^{\mathrm{c}}\left|\M_\NC
-\left[{\cal M}_{\PW}
\right]_{\{p_1,Q_1,\si_1\} \lra \{p_3,Q_3,\si_3\}}\right|^2,
\eeqar
where we use the shorthand
\beq
\M_\NC = \sum_{\PV=\gamma,\PZ}{\cal M}_{\PV}
\eeq
and suppress the helicity indices and the dependence on 
momenta and relative charges.
The relative signs account for interchanging external fermion lines.

\subsubsection{Squared amplitudes for hadronic final states}
\label{se:hadfinstat}

Next we consider purely hadronic final states, i.e., the cases where
all final-state fermions are quarks. 
This renders the summation
of the squared matrix elements over the colour degrees of freedom 
non-trivial, and in addition gluon-exchange diagrams appear.
Since gluon-exchange diagrams require two quark--anti-quark
pairs in the final state they do not appear in CC processes. 
For CC processes there is only one possibility for the colour flow,
and the summation over the colour degrees of freedom 
leads to an overall factor 
$N^{\mathrm{c}}_{\mathrm{had,CC}}=3^2=9$ to the squared matrix elements
as given in Eq.~\refeq{eq:MCC2}.

For NC reactions we have to compute the sum of pure electroweak 
(ew) and 
gluon-exchange (QCD) matrix elements,
\beq
\M_{\had}^{c_1c_2c_3c_4} = \M_{\had,\ew}^{c_1c_2c_3c_4} 
+ \M_{\had,\QCD}^{c_1c_2c_3c_4},
\eeq
where $c_i$ denotes the colour indices of the quarks.
The electroweak diagrams are diagonal in colour space and 
therefore read
\beqar
\M_{\NC(\mathrm{a}),\had,\ew}^{c_1c_2c_3c_4}
&=& \M_{\NC}\delta_{c_1c_2}\delta_{c_3c_4}, 
\nn \\
\M_{\NC(\mathrm{b}),\had,\ew}^{c_1c_2c_3c_4}
&=& \M_{\NC}\delta_{c_1c_2}\delta_{c_3c_4}
- \left[\M_{\NC}\right]_{\{p_1,Q_1,\si_1\} \lra \{p_3,Q_3,\si_3\}} \,
\delta_{c_3c_2}\delta_{c_1c_4}.
\eeqar
The gluon-exchange diagrams are obtained from the generic 
formula \refeq{genfun4f} by inserting 
the corresponding generators, $\la^a/2$, of the gauge group $\SU(3)$,
\beqar
\M_{\NC(\mathrm{a}),\had,\QCD}^{c_1c_2c_3c_4}
&=& \M_{\Pg}\frac{1}{4}\lambda^a_{c_1c_2}\lambda^a_{c_3c_4},
\nn \\
\M_{\NC(\mathrm{b}),\had,\QCD}^{c_1c_2c_3c_4}&=&
\M_{\Pg}\frac{1}{4}\lambda^a_{c_1c_2}\lambda^a_{c_3c_4}
- \left[\M_{\Pg}\right]_{\{p_1,Q_1,\si_1\} \lra \{p_3,Q_3,\si_3\}} \,
\frac{1}{4}\lambda^a_{c_3c_2}\lambda^a_{c_1c_4}.
\eeqar
The matrix element $\M_{\Pg}$ is defined by 
Eq.~\refeq{genfun4f} with $V=\Pg$. 

Carrying out the colour sum using the completeness relation 
for the Gell-Mann matrices,
\beq
\lambda_{ij}^a\lambda_{kl}^a=-\frac{2}{3}\delta_{ij}\delta_{kl}+
2\delta_{il}\delta_{jk},
\eeq
yields
\beqar
\label{hadronic}
\lefteqn{\sum_{\mathrm{colour}}|\M_{\NC(\mathrm{a}),\had}|^2 = 
9|\M_{\NC}|^2+2|\M_{\Pg}|^2,}&&
\nn\\ 
\lefteqn{\sum_{\mathrm{colour}}|\M_{\NC(\mathrm{b}),\had}|^2 = 
9|\M_{\NC}|^2
+9\left|\left[\M_{\NC}
\right]_{\{p_1,Q_1,\si_1\} \lra \{p_3,Q_3,\si_3\}}\right|^2
+2|\M_{\Pg}|^2 }
&&
\nn\\
&& \qquad {}
+2\left|\left[\M_{\Pg}\right]_{\{p_1,Q_1,\si_1\} \lra \{p_3,Q_3,\si_3\}}\right|^2
-6\Re\left\{\M_{\NC} \left[\M^*_{\NC}
\right]_{\{p_1,Q_1,\si_1\} \lra \{p_3,Q_3,\si_3\}}\right\}
\nn\\
&& \qquad {}
+\frac{4}{3}\Re\left\{
\M_{\Pg}\left[\M^*_{\Pg}\right]_{\{p_1,Q_1,\si_1\} \lra \{p_3,Q_3,\si_3\}}\right\} 
-8\Re\left\{\M_{\NC}\left[\M^*_{\Pg}\right]_{\{p_1,Q_1,\si_1\} \lra \{p_3,Q_3,\si_3\}}\right\}
\nn\\
&& \qquad {}
-8\Re\left\{\M_{\Pg}\left[\M^*_{\NC}\right]_{\{p_1,Q_1,\si_1
\} \lra \{p_3,Q_3,\si_3\}}\right\}.
\eeqar

All squared matrix elements of this section have been numerically compared 
with results obtained with the program {\Madgraph} \cite{Stelzer:1994ta} 
at several phase-space points, and perfect agreement has been found.

\subsection{Implementation of finite gauge-boson widths}
\label{se:finwidth}

\begin{sloppypar}
We have implemented the finite widths of the W- and Z-boson 
propagators%
\footnote{We have also supplemented the explicit gauge-boson
masses appearing in the numerators of Eq.~\refeq{eq:24ampl} by the
corresponding widths, because these mass terms originate from
denominators upon combining different diagrams.}
in four different ways:
\begin{itemize}
\item{\it fixed width} in all propagators:
\beq
P_{\PV}(p) = \frac{1}{p^2-\MV^2+\ri\MV\GV},
\eeq
\item{\it step width} (fixed width in time-like propagators): 
\beq
P_{\PV}(p) = \frac{1}{p^2-\MV^2+\ri\MV\GV\theta(p^2)},
\eeq
\item{\it running width} in time-like propagators: 
\beq
P_{\PV}(p) = \frac{1}{p^2-\MV^2+\ri p^2(\GV/\MV)\theta(p^2)},
\eeq
\item{\it complex-mass scheme} \cite{Denner:1999gp}:
complex gauge-boson masses are used everywhere,
\ie $\sqrt{\MV^2-\ri \MV\Ga_V}$ instead of $\MV$ in all propagators
and couplings. This results in a constant width in all propagators,
\beq\label{Vprop}
P_{\PV}(p) = \frac{1}{p^2-\MV^2+\ri\MV\Ga_{\PV}},
\eeq
and in a complex weak mixing angle 
\beq\label{complangle}
\cw^2=1-\sw^2=   \frac{\MW^2-\ri\MW\GW}{\MZ^2-\ri\MZ\GZ}.
\eeq
\end{itemize}
\end{sloppypar}

The virtues and drawbacks of the first three schemes are discussed 
in \citere{Argyres:1995ym}. All but the complex-mass scheme, in general,
violate $\SU(2)$ gauge invariance, the step- and the 
running-width schemes also violate electromagnetic $\U(1)_{\mathrm{em}}$ 
gauge invariance, which is preserved by using a fixed width.
As known from many examples in $\Pep\Pem$ physics
\cite{Denner:1999gp,Argyres:1995ym,Dittmaier:2002ap},
gauge-invariance-violating effects, in particular when enhanced by
factors $p^2/\MV^2$ as in the running-width scheme, 
can lead to totally wrong results. 
Furthermore, the violation of $\U(1)_{\mathrm{em}}$ gauge 
invariance also causes a dependence of matrix elements and
cross sections on the gauge chosen for external photons. 
In $\eeffff$ and $\eeffffff$ this problem does not occur since no 
external photons are involved.

The complex-mass scheme, which was introduced in \citere{Denner:1999gp}
for tree-level calculations,
preserves gauge invariance 
and thus all Ward identities which rule gauge cancellations.
Its application is particularly simple for $\ggffffkg$ in the
non-linear gauge. In this case, no couplings
involving explicit gauge-boson masses appear,
and it is sufficient to
introduce the finite gauge-boson widths in the propagators [\cf
Eq.~\refeq{Vprop}] and to introduce the complex weak mixing angle
\refeq{complangle} in the couplings. 

For CC processes $\ggffff(\ga)$ 
with massless fermions,
the fixed-width (FW) approach in the non-linear gauge
and the complex-mass scheme (CMS) are practically equivalent, because
all Feynman graphs are proportional to $e^4/\sw^2$ 
($e^5/\sw^2$) and gauge-boson
masses appear only in propagator denominators. In this case
the corresponding amplitudes in the two schemes
differ only in the global factor
$s_{\Pw,\mathrm{FW}}^2 / s_{\Pw,\mathrm{CMS}}^2$, 
where $s_{\Pw,\mathrm{FW}}$ and $s_{\Pw,\mathrm{CMS}}$ are the
values of $\sw$ in the different schemes,
i.e., $s_{\Pw,\mathrm{FW}}$ is derived from the ratio of real
gauge-boson masses and $s_{\Pw,\mathrm{CMS}}$ from complex masses.
Thus, both squared amplitudes are gauge invariant and are equal up to 
the factor $|s_{\Pw,\mathrm{FW}} / s_{\Pw,\mathrm{CMS}}|^4$ which is 
equal to 1 up to terms of ${\cal O}(\GW^2/\MW^2)$.

For NC and CC/NC processes a similar reasoning can be used to
show that the fixed-width approach does not violate gauge
invariance in $\ggffff(\ga)$ for massless fermions.
The trick is to apply the above argument to gauge-invariant 
subsets of diagrams.
For NC diagrams with photon exchange, which is the (gauge-invariant)
QED subset of diagrams (\reffi{fig:gengraph} with $V=\ga$), 
there is nothing to show. The sum of NC diagrams of type NC(a)
with Z-boson exchange (\reffi{fig:gengraph} with $V=\PZ$) again
involves $\cw$ and $\sw$ only in a global coupling factor 
(per helicity channel); the remaining
dependence on the gauge-boson masses is located in the propagator
denominators. Thus, the subamplitudes of the fixed-width and the complex-mass
scheme are again identical up to a global factor and both preserve
gauge invariance and Ward identities. For NC processes of type
NC(b) a second class of diagrams exists (\reffi{fig:gengraph} with 
$V=\ga,\PZ$ and external fermions interchanged). This new class of diagrams 
forms a gauge-invariant subset because of the different flow of fermion 
numbers. Thus, the 
reasoning for type NC(a) applies to both classes of diagrams of
NC(b) reactions. The same argument is also valid for the subset
of CC diagrams in mixed CC/NC reactions. 

In summary, we have argued that the use of naive fixed
gauge-boson widths does not lead to gauge-invariance violations
in amplitudes for $\ggffff(\ga)$ as long as fermions are massless
and the non-linear gauge with vanishing $\ga\PW\phi$ coupling
(or the complex W-boson mass in this coupling if the
`t~Hooft--Feynman gauge is chosen) is used.
The corresponding squared amplitudes agree with the ones of the
(gauge-invariant) complex-mass scheme up to terms of
${\cal O}(\GW/\MW)$, for CC processes even up to terms of
${\cal O}(\GW^2/\MW^2)$.

\subsection{W-pair signal diagrams and double-pole approximation}
\label{se:dpa}

The diagrams to CC and CC/NC processes comprise graphs with two,
one, or no internal W-boson lines that can become resonant,
similar to the situation for $\Pep\Pem\to\PW\PW\to4f$
(see \citeres{Grunewald:2000ju,Beenakker:1996kt} and references
therein).
It is interesting to investigate the possibility to define an
amplitude for the W-pair signal based on doubly-resonant contributions
only, because such an amplitude is much simpler than the full
amplitudes for four-fermion production and is universal (up to colour factors)
for all relevant $4f$ final states.
Moreover, this study is an important exercise for the calculation of 
radiative corrections to $\ggwwffff$ in the so-called 
double-pole approximation (DPA),
where only doubly-resonant contributions are taken into account.
Taking simply all doubly-resonant diagrams, of course, yields
a result that is not gauge invariant. Nevertheless in the $\Pep\Pem$
case the lowest-order cross section based on such a gauge-dependent amplitude
(defined in the `t~Hooft--Feynman gauge),
known as ``CC03 cross section'', is a very useful quantity that
is very close to the full $4f$ calculation if 
both W~bosons are close to resonance.
The CC03 amplitude can be rendered gauge invariant upon deforming the
momenta of the four outgoing fermions in such a way that the
intermediate W-boson states become on shell, because the
residues of the W~resonances are gauge-invariant quantities.
This ``on-shell projection'', which is part of the 
DPA, involves some freedom, and different versions, which
have been described in \citeres{Beenakker:1998gr,Denner:2000kn},
differ by contributions of relative order ${\cal O}(\GW/\MW)$.

We want to perform the exercise to study the usefulness of a possible
``CC03''%
\footnote{The name also fits to the $\ga\ga$ case where three
W-pair diagrams exist in unitary or non-linear gauge.}
off-shell cross section for $\ggwwffff$.
To this end, we define the amplitude for the off-shell W-pair signal
by evaluating the three W-pair diagrams in the non-linear gauge
with polarization vectors $\veps_i(k_i)$ for the incoming photons,
which obey the gauge conditions 
\beq
\veps_1(k_1)\cdot k_2 = \veps_2(k_2)\cdot k_1 = 0.
\eeq
In terms of WvdW 
spinors, this means that the gauge spinors $g_1$ and $g_2$ of the photons 
are identified with the 
spinors of the momenta $k_2$ and $k_1$, respectively.
With this choice the auxiliary functions for the matrix elements 
\refeq{genfun4f} read 
\beqar
\lefteqn{A^{--}_{++,\PW\PW}(k_i,p_j,Q_j)=P_{\PW}(p_1+p_2)P_{\PW}(p_3+p_4)
\frac{\cspdf}{\cspab}}
&&
\nn\\
&& {}
\times
\Biggl\{
\Big[P_{\PW}(p_1+p_2-k_1)\Big]_{\GW=0}
\Biggl[ 
\spbc\spbe\langle k_2[P_1+P_2]k_1\rangle
+\spac\spae\langle k_1[P_3+P_4]k_2\rangle
\nn\\
&& \qquad\qquad {}
+\frac{\spce}{\cspab}\langle k_2[P_1+P_2]k_1\rangle
\langle k_1[P_3+P_4]k_2\rangle
-2(k_1\cdot k_2)\spac\spbe
\Biggr]
\nn\\
&& \qquad {}
-\frac{1}{2}\spce\spab\Biggr\}
+(k_1\lra k_2),
\nn\\
\lefteqn{A^{--}_{+-,\PW\PW}(k_i,p_j,Q_j)=P_{\PW}(p_1+p_2)P_{\PW}(p_3+p_4)
} &&
\nn\\
&& {}
\times
\Big[P_\PW(p_1+p_2-k_1)+P_\PW(p_1+p_2-k_2)\Big]_{\GW=0}
\nn\\
&& {}
\times \Biggl\{
\langle k_2[P_1+P_2]k_1\rangle \Biggl[
\frac{\cspbd\cspbf\spce}{\cspab}
-\frac{\cspdf\spac\spae}{\spab}
\Biggr]
\nn\\
&& \qquad {}
-\frac{\cspdf\spce}{2(k_1 k_2)}
\langle k_2[P_1+P_2]k_1\rangle^2
+\cspbd\cspbf\spac\spae
\Biggr\}.
\label{eq:WWsignal}
\eeqar
Note that $A^{\si_1\si_3}_{\la_1\la_2,\PW\PW}$ do not 
coincide with the parts of the functions $A^{\si_1\si_3}_{\la_1\la_2,\PW}$ 
of Eq.~\refeq{eq:24ampl} 
that are proportional to $P_\PW(p_1+p_2)P_\PW(p_3+p_4)$
because the derivation of Eq.~\refeq{eq:24ampl} 
involves rearrangements of various singly-resonant contributions.
We point out that the definition \refeq{eq:WWsignal}
is neither independent of the gauge fixing 
used to define gauge-boson propagators nor of the gauge
of the external photons.
The definition is gauge invariant after the outgoing fermion
momenta $p_i$ are on-shell projected as described above, while
leaving the resonant propagators
$P_{\PW}(p_1+p_2)P_{\PW}(p_3+p_4)$ untouched.
This defines the lowest-order amplitude in DPA.
Finally, we stress that the $t$- and $u$-channel W~propagators 
in Eq.~\refeq{eq:WWsignal} do not receive a finite W~width;
otherwise the gauge invariance of the DPA would be spoiled.

\section{Inclusion of anomalous gauge-boson couplings}
\label{se:agc}

In this section we introduce the most important anomalous gauge-boson
couplings accessible by the process $\ggffff$ and give explicit
analytical results for the corresponding helicity amplitudes.

\subsection{The effective Lagrangians}
\label{se:efflagr}

First we consider anomalous triple gauge-boson couplings (ATGC) in the 
charged-current sector, i.e., 
anomalous $\gww$ and the related $\ggww$ couplings.
Instead of using rather general parametrizations of non-standard
couplings \cite{Gaemers:1978hg}, we follow the approach already
used at LEP2 to reduce the number of free parameters by
requiring that all symmetries of the SM are respected.
From the resulting operators we only keep those that appear in the 
lowest-order cross section of $\ggffff$.
Specifically, we start from the gauge-invariant CP-conserving 
effective Lagrangian with dimension-6 operators \cite{Gounaris:1996rz}%
\beqar
\label{eq:Ltriple}
{\cal L}^{\mathrm{ATGC}}_{\CC} &=&  \ri g_1 {\frac{\aBphi}{\MW^2}}\,
 (D_\mu\Phi)^\dagger B^{\mu\nu}(D_\nu\Phi)\,
-\, \ri g_2 {\frac{\aWphi}{\MW^2}} \,
(D_\mu\Phi)^\dagger {\mbox{\boldmath$\si$}} \cdot {\bf{W}}^{\mu\nu}
(D_\nu\Phi)\,
\nn\\
&&-g_2 {\frac{\aW}{6\MW^2}}\,
{\bf{W}}^{\mu}_{\phantom{\mu}\nu}
\cdot ({\bf{W}}^\nu_{\phantom{\nu}\rho}
\times {\bf{W}}^{\rho}_{\phantom{\rho}\mu}),
\eeqar
where $\Phi$ is the Higgs doublet field and
\beqar
B^{\mu\nu} &=& \partial^\mu B^\nu-\partial^\nu B^\mu,
\nn\\
{\bf W}^{\mu\nu} &=& (W_1^{\mu\nu},W_2^{\mu\nu},W_3^{\mu\nu})
= \partial^\mu {\bf{W}}^\nu-\partial^\nu {\bf{W}}^\mu
+g_2 {\bf{W}}^\mu \times {\bf{W}}^\nu
\eeqar
are the field strengths of the U(1) and SU(2) gauge fields, respectively. 
The Pauli matrices are combined into the vector 
$\mbox{\boldmath$\si$}=(\si_1,\si_2,\si_3)$, and
the parameters $g_1$, $g_2$ denote the 
gauge couplings.%
\footnote{In order to be compatible with the conventions of 
\citere{Bohm:rj} used for the SM amplitudes above, we had to change the
sign of the SU(2) coupling $g_2$ w.r.t.\ \citere{Gounaris:1996rz}.}
Inserting the vacuum expectation value 
of the Higgs field $\Phi$, we can relate the coefficients 
$\aBphi$, $\aWphi$, and $\aW$ to the coefficients of the Lagrangian 
considered in the LEP2 analysis
\cite{Gounaris:1996rz},
\beq
\label{tgcd6}
\Delta g_1^{\PZ} = \frac{\aWphi}{\cw^2}, \qquad
\Delta\kappa_\gamma =  -\frac{\cw^2}{\sw^2}(
\Delta\kappa_{\PZ} -\Delta g_1^{\PZ} ) =  \aWphi +\aBphi, \qquad
\lambda_\gamma = \lambda_{\PZ}  =  \aW.
\eeq
In contrast to the pure anomalous $\gww$ coupling 
\cite{Gaemers:1978hg}, the SU(2)$\times$U(1) symmetry of the effective
Lagrangian \refeq{eq:Ltriple} induces additional
anomalous $\ggww$ and 
$\ga\PW\phi$ couplings.
The corresponding Feynman rules are
\beqar
\lefteqn{\ri\Gamma^{\gamma \PW^+\PW^-}_{\mu\nu\rho}(k_0,k_+,k_-)=
-\ri e\Biggl\{\Delta\ka_{\ga}(k_{0\nu}g_{\mu\rho}
-k_{0\rho} g_{\mu\nu})} &&
\nn \\
&& \qquad {}
-\frac{\lambda_{\ga}}{\MW^2}
\Biggl[k_{+\mu}k_{-\nu}k_{0\rho}-k_{-\mu}k_{+\rho}k_{0\nu}
+g_{\nu\rho}(k_{-\mu}(k_+k_0)-k_{+\mu}(k_-k_0))
\nn \\
&& \qquad\qquad {}
+g_{\mu\rho}(k_{0\nu}(k_+k_-)-k_{-\nu}(k_+k_0))
+g_{\mu\nu}(
k_{+\rho}(k_-k_0)
-k_{0\rho}(k_+k_-)
)\biggr]\Biggr\},
\nn\\[.5em]
\lefteqn{\ri\Gamma^{\ga\ga \PW^+\PW^-}_{\mu\nu\rho\si}(k_1,k_2,k_+,k_-)
= -\ri e^2\frac{\la_{\ga}}{\MW^2}\Biggl\{
g_{\mu\nu}g_{\rho\si}(k_1+k_2)^2
+g_{\mu\rho}g_{\nu\si}(k_2k_++k_1k_-) 
} && 
\nn \\
&& \qquad {}
+ g_{\nu\rho}g_{\mu\si}(k_1k_++k_2k_-)
+g_{\mu\nu}\Bigl[(k_1+k_2)_{\rho}k_{+\si}+(k_1+k_2)_{\si}k_{-\rho}\Bigr]
\nn \\
&& \qquad {}
+g_{\rho\si}\Bigl[(k_++k_-)_{\mu}k_{1\nu}+(k_++k_-)_{\nu}k_{2\mu}\Bigr]
+g_{\mu\rho}\Bigl[(k_1-k_2)_{\si}k_{+\nu}-k_{1\nu}k_{+\si}
-k_{1\si}k_{-\nu}\Bigr]
\nn \\
&& \qquad {}
+g_{\mu\si}\Bigl[(k_1-k_2)_{\rho}k_{-\nu}-k_{1\nu}k_{-\rho}-k_{1\rho}k_{+\nu}\Bigr]
+g_{\nu\rho}\Bigl[(k_2-k_1)_{\si}k_{+\mu}-k_{2\mu}k_{+\si}
-k_{2\si}k_{-\mu}\Bigr]
\nn \\
&& \qquad {}
+g_{\nu\si}\Bigl[(k_2-k_1)_{\rho}k_{-\mu}-k_{2\mu}k_{-\rho}-k_{2\rho}k_{+\mu}\Bigr]\Biggr\},
\nn\\[.5em]
\lefteqn{\ri\Gamma^{\ga\PW\phi}_{\mu\nu}(k_0,k_{\PW},k_{\phi})=
-\ri e\frac{\Delta\ka_\ga}{\MW}\biggl\{(k_{\phi}k_0)g_{\mu\nu}-
k_{\phi,\mu}k_{0,\nu}\biggr\},}
\eeqar
where all fields and momenta are considered incoming.
Note that the neglect of the contribution to the quartic coupling 
$\ga\ga\PW\PW$, which is proportional to $\la_\ga$, would lead to
a violation of electromagnetic gauge invariance in predictions for
$\ga\ga\to\PW\PW(\to 4f)$. 
In contrast, neglecting the $\ga\PW\phi$ coupling, which is proportional 
to $\De\kappa_\ga$, would not spoil the electromagnetic gauge invariance
of the predictions.

Next we consider anomalous triple gauge-boson couplings involving only 
the neutral gauge bosons $\ga$ and Z.
Assuming Lorentz invariance and electromagnetic gauge invariance, the most
general effective dimension-6 Lagrangian for $\ggz$, 
$\gzz$, and $\zzz$ couplings can be written as 
\cite{Gounaris:2000kf}%
\footnote{Note that our conventions differ from those of
\citere{Gounaris:2000kf} by a minus sign in the Z-boson field.}
\beqar \label{eq:Lncatgc}
{\cal{L}}^{\mathrm{ATGC}}_{\NC} &=& \frac{e}{\MZ^2}\biggl\{
[f_4^\ga(\partial_\mu F^{\mu\nu})-f_4^\PZ(\partial_\mu Z^{\mu\nu})]
Z_{\nu\rho}Z^\rho
+[f_5^\ga(\partial_\mu F^{\mu\nu})-f_5^\PZ(\partial_\mu Z^{\mu\nu})]
\tilde Z_{\nu\rho}Z^\rho\nl
&&{}
+[h_1^\ga(\partial_\mu F^{\mu\nu})-h_1^\PZ(\partial_\mu Z^{\mu\nu})]
F_{\nu\rho}Z^\rho
+[h_3^\ga(\partial_\mu F^{\mu\nu})-h_3^\PZ(\partial_\mu Z^{\mu\nu})]
\tilde F_{\nu\rho}Z^\rho\biggr\}
\eeqar
with the abelian field-strength tensors
\beq
F^{\mu\nu}=\partial^{\mu}A^{\nu}-\partial^{\nu}A^{\mu},\qquad
Z^{\mu\nu}=\partial^{\mu}Z^{\nu}-\partial^{\nu}Z^{\mu}
\eeq
and the dual field-strength tensors ($\eps^{0123}=+1$)
\beq 
\tilde F^{\mu\nu}=\frac{1}{2}\eps^{\mu\nu\rho\si}F_{\rho\si},\qquad
\tilde Z^{\mu\nu}=\frac{1}{2}\eps^{\mu\nu\rho\si}Z_{\rho\si}.
\eeq 
An operator inducing a $\ga\ga\ga$ coupling does not appear in 
Eq.~\refeq{eq:Lncatgc} since it violates electromagnetic 
gauge invariance.

Apart from the $\ggww$ coupling which is induced by symmetries in 
the Lagrangian \refeq{eq:Ltriple},
we also include genuine anomalous quartic gauge-boson couplings (AQGC)
in our analysis, 
whose lowest dimension is 6.
In \citeres{Belanger:1992qh,Denner:2001vr} all genuine dimension-6 AQGC
that involve photons and that are allowed by electromagnetic 
gauge invariance and custodial $\SU(2)_{\mathrm{c}}$ 
have been classified;
more general AQGC have been discussed in \citere{Belanger:2000aw}.
Following \citere{Denner:2001vr} we use the effective Lagrangian
\newcommand{\oW}{\overline{W}}
\newcommand{\boW}{\mathbf{\oW}}
\beq
{\cal L}^{\mathrm{AQGC}}_{\ga\ga\PV\PV} =  
-\frac{e^2}{16\La^2} \biggl\{
\az\, F^{\mu\nu} F_{\mu\nu} \boW_{\!\al} \boW^\al
+\ac\, F^{\mu\al} F_{\mu\be} \boW^\be \boW_{\!\al}
+\azt\,  F^{\mu\nu} \tilde F_{\mu\nu} \boW_{\!\al} \boW^\al
\biggr\}
\label{eq:QGCs}
\eeq
with the definition
\beq
{\boW}_{\!\mu} = \left(\oW^1_\mu,\oW^2_\mu,\oW^3_\mu\right) =
\left(\frac{1}{\sqrt{2}}(W^++W^-)_\mu,\frac{\ri}{\sqrt{2}}(W^+-W^-)_\mu,
\frac{1}{\cw}Z_\mu\right).
\eeq
The scale of new physics, 
$\La$, is introduced in Eq.~\refeq{eq:QGCs} to
render the coupling coefficients $a_0,a_{\mathrm{c}},\tilde{a}_0$ 
dimensionless.
The effective Lagrangian ${\cal L}^{\mathrm{AQGC}}_{\ggvv}$
contains $\ggww$ and $\ggzz$ couplings, whose
Feynman rules can be found in \citere{Denner:2001vr}.
The other coupling structures 
$\L_{\mathrm{n}}$ and $\tilde\L_{\mathrm{n}}$ 
considered in \citere{Denner:2001vr}
induce $\gzww$ couplings that are not relevant for
$\ggffff$.

\subsection{Amplitudes with triple gauge-boson couplings}

\bfi
\centerline{
\begin{picture}(120,110)(0,0)
\Text(0,90)[lb]{(a)}
\Photon(12,70)(45,45){2}{4}
\Photon(12,20)(45,45){2}{4}
\GCirc(45,45){5}{0}
\Photon(45,45)(80,70){2}{4}
\Photon(45,45)(80,20){2}{4}
\Vertex(80,70){2}
\ArrowLine(80,70)(100,80)
\ArrowLine(100,60)(80,70)
\Vertex(80,20){2}
\ArrowLine(80,20)(100,30)
\ArrowLine(100,10)(80,20)
\put(-5,70){$\ga_1$}
\put(-5,20){$\ga_2$}
\put(52,64){$\PW$}
\put(52,18){$\PW$}
\put(105,78){$f_1$}
\put(105,58){$\bar f_2$}
\put(105,28){$f_3$}
\put(105,8){$\bar f_4$}
\end{picture}
\hspace*{2em}
\begin{picture}(120,110)(0,0)
\Text(0,90)[lb]{(b)}
\Photon(15,70)(50,70){2}{4}
\Photon(15,20)(50,20){2}{4}
\Vertex(50,70){5}
\Vertex(50,20){2}
\Photon(50,70)(50,20){2}{6}
\Photon(50,70)(90,70){2}{5}
\Photon(50,20)(90,20){2}{5}
\Vertex(90,70){2}
\ArrowLine(90,70)(110,80)
\ArrowLine(110,60)(90,70)
\Vertex(90,20){2}
\ArrowLine(90,20)(110,30)
\ArrowLine(110,10)(90,20)
\put(0,70){$\ga_1$}
\put(0,20){$\ga_2$}
\put(65,75){$\PW$}
\put(55,40){$\PW$}
\put(65,5){$\PW$}
\put(115,78){$f_1$}
\put(115,58){$\bar f_2$}
\put(115,28){$f_3$}
\put(115,8){$\bar f_4$}
\end{picture}
\hspace*{2em}
\begin{picture}(120,110)(0,0)
\Text(0,90)[lb]{(c)}
\ArrowLine(30,45)(10,65)
\ArrowLine(10,25)(30,45)
\put(-7,63){$f_1$}
\put(-7,23){$\bar f_2$}
\Photon(30,45)(80,45){2}{6}
\Vertex(55,45){5}
\Photon(55,45)(55,20){2}{4}
\put(50,10){$\ga_1$}
\ArrowLine(80,45)(90,55)
\ArrowLine(90,55)(100,65)
\ArrowLine(100,25)(80,45)
\Photon(90,55)(110,55){2}{3}
\put(105,63){$f_3$}
\put(105,23){$\bar f_4$}
\put(113,52){$\ga_2$}
\put(35,52){$\PW$}
\put(62,52){$\PW$}
\end{picture}}
\vspace{1em}
\centerline{
\begin{picture}(120,110)(0,0)
\Text(0,90)[lb]{(d)}
\Photon(15,70)(45,70){2}{4}
\Photon(15,20)(45,20){2}{4}
\Vertex(50,70){5}
\Vertex(50,20){5}
\Photon(50,70)(50,20){2}{6}
\Photon(50,70)(90,70){2}{5}
\Photon(50,20)(90,20){2}{5}
\Vertex(90,70){2}
\ArrowLine(90,70)(110,80)
\ArrowLine(110,60)(90,70)
\Vertex(90,20){2}
\ArrowLine(90,20)(110,30)
\ArrowLine(110,10)(90,20)
\put(0,70){$\ga_1$}
\put(0,20){$\ga_2$}
\put(65,75){$\PW$}
\put(55,40){$\PW$}
\put(65,5){$\PW$}
\put(115,78){$f_1$}
\put(115,58){$\bar f_2$}
\put(115,28){$f_3$}
\put(115,8){$\bar f_4$}
\end{picture}
\hspace*{2em}
\begin{picture}(120,110)(0,0)
\Text(0,90)[lb]{(e)}
\Photon(15,70)(45,70){2}{4}
\Photon(15,20)(45,20){2}{4}
\Vertex(50,70){5}
\Vertex(50,20){5}
\DashLine(50,70)(50,20){5}
\Photon(50,70)(90,70){2}{5}
\Photon(50,20)(90,20){2}{5}
\Vertex(90,70){2}
\ArrowLine(90,70)(110,80)
\ArrowLine(110,60)(90,70)
\Vertex(90,20){2}
\ArrowLine(90,20)(110,30)
\ArrowLine(110,10)(90,20)
\put(0,70){$\ga_1$}
\put(0,20){$\ga_2$}
\put(65,75){$\PW$}
\put(55,40){$\phi$}
\put(65,5){$\PW$}
\put(115,78){$f_1$}
\put(115,58){$\bar f_2$}
\put(115,28){$f_3$}
\put(115,8){$\bar f_4$}
\end{picture}}
\caption{Representative diagrams with anomalous $\ga\PW\PW$ and $\ga\ga\PW\PW$
couplings (black blobs) contributing to CC processes $\ggffff$.}
\label{fig:ATGC}
\efi
Before we write down the helicity amplitudes including ATGC explicitly, 
we discuss the impact of these couplings w.r.t.\ the SM cross section. 
The diagrams containing ATGC and the corresponding quartic couplings
in CC diagrams are shown in \reffi{fig:ATGC}.
We quantify the size of the anomalous contributions
in terms of powers of anomalous coupling factors (generically
denoted by $a_3$) or suppression factors $\GW/\MW$.
Considering the SM process $\ggwwffff$ as the leading
contribution, i.e., regarding anomalous-coupling effects as small,
we get non-standard contributions to CC and CC/NC cross sections
from CC ATGC of the following orders:
\begin{itemize}
\item 
$\Ord(a_3)$:\\
The matrix elements of diagrams (a) and (b) in 
\reffi{fig:ATGC} involve one power of $a_3$. Both diagrams are 
not suppressed by $\GW/\MW$ since they are doubly resonant.
\item
$\Ord(a_3\GW/\MW)$:\\
The diagram (c) of \reffi{fig:ATGC} has one power of $a_3$ and
one resonant W-boson propagator, i.e., it is only 
singly resonant. Thus, it is of $\Ord(a_3\GW/\MW)$.
\item
$\Ord(a_3^2)$:\\
The diagrams (d) and (e) of \reffi{fig:ATGC} involve two anomalous 
couplings $a_3$ and are doubly resonant.
Therefore, they are of 
$\Ord(a_3^2)$. Note that the squares of the diagrams (a) and (b),
as well as their products with one another,
are of the same order
as the interference of diagrams (d) and (e) with the SM amplitude.
\end{itemize}
There are no diagrams containing CC ATGC for NC processes.

Next we consider the impact of NC ATGC, as defined in the effective 
Lagrangian \refeq{eq:Lncatgc}. 
The by far largest SM cross sections of the process class $\ggffff$ belong
to diagrams with two resonant W bosons in CC and CC/NC reactions. 
Thus, the largest effect of NC ATGC could be expected from
an interference of ``anomalous diagrams'' with the SM amplitude for
CC or CC/NC processes. The only candidate of this kind is a diagram where
an off-shell $s$-channel Z~boson is produced by an anomalous 
$\ggz$ coupling that subsequently produces a W-boson pair. 
However, the effective $\ggz$ coupling of Eq.~\refeq{eq:Lncatgc}
vanishes for two on-shell photons, so that this diagram does
not contribute. No other CC diagram exists that includes 
a NC ATGC.

We now turn to the effects of NC ATGC in NC amplitudes, i.e., in
diagrams without W~bosons. The corresponding SM
amplitudes involve at most a single resonance of the Z~boson, 
which leads already to a suppression of NC cross sections w.r.t.\ CC
cross sections by a factor $(\GZ/\MZ)^2$. This
suppression is clearly visible in the numerical results presented 
in \refse{se:cross} below.
Diagrams with one NC ATGC also possess at most one resonant Z~boson
and, therefore, show a suppression by a factor $a_3(\GZ/\MZ)^2$
w.r.t.\ the CC signal diagrams. 
This suppression is not changed
by interferences with doubly-resonant CC diagrams in CC/NC processes
because the Z- and W-boson resonances are located at different
regions in phase space and do not enhance each other.
Diagrams with two NC ATGC can involve two Z-boson 
resonances resulting in a suppression of ${\cal O}(a_3^2\GZ/\MZ)$, which is
also small compared to the CC case owing to the squared ATGC.
In summary, we conclude that the sensitivity of the processes
$\ggffff$ to NC ATGC is much smaller than to CC ATGC. Therefore,
we restrict our investigation on ATGC to CC couplings in the following.

As explained above, the diagrams of \reffi{fig:ATGC} induce contributions
to the amplitude that are either linear or quadratic in the CC ATGC.
We give the explicit contributions to the helicity amplitudes in
a way similar to the SM case \refeq{genfun4f},
\beq\label{genfunATGC}
\M_{\la_1\la_2,\mathrm{CCATGC}}^{\si_1\si_2\si_3\si_4}
(k_i,p_j,Q_j) 
= e^4 \de_{\si_1,-} \de_{\si_2,+} \de_{\si_3,-} \de_{\si_4,+} \, 
g^-_{\PW\bar f_1 f_2} g^-_{\PW\bar f_3 f_4} \,
\de_3A^{\si_1\si_3}_{\la_1\la_2}(k_i,p_j,Q_j)
\eeq
with the auxiliary functions 
$\de_3A^{\si_1\si_3}_{\la_1\la_2}$.
The generic amplitude $\M_{\la_1\la_2,\mathrm{CCATGC}}^{\si_1\si_2\si_3\si_4}$
is coherently added to the SM amplitude 
$\M_{\la_1\la_2,\PW}^{\si_1\si_2\si_3\si_4}$ of
Eq.~\refeq{genfun4f}. The colour summation of the squared amplitudes for
the various process types proceeds as described in
\refses{se:lepfinstat} and \ref{se:hadfinstat}.

The terms in $\de_3A^{\si_1\si_3}_{\la_1\la_2}$ that are quadratic and linear
in ATGC explicitly read
\beqar
\lefteqn{\de_3A^{--}_{++}\Big|_{\mathrm{quad}} = 
- P_{\PW}(p_1+p_2)P_{\PW}(p_3+p_4)P_{\PW}(p_1+p_2-k_1)\spac\spbe} &&
\nn\\
&& {}
\times \left\{\dkag^2\biggl[\cspdf\spab
+\frac{1}{2\MW^2}\cspcd\cspef\spac\spbe\biggr]\right.
\nn\\
&& \qquad {}
+\dkag\frac{\lag}{\MW^2}\biggl[\cspcd\cspef\Bigl(
\spae\spbc-\spab\spce\Bigr)
\nn\\
&& \qquad {}
+\spab\Bigl(\cspef\cspad\spae -\cspcd\cspbf\spbc\Bigr)\biggr]
\nn\\
&& \qquad {}
\left.+\frac{\lag^2}{\MW^4}\cspcd\cspef\frac{1}{2}(p_1+p_2-k_1)^2
\biggl[\spae\spbc-\spab\spce\biggr]\right\}
\nn\\*
&& {}
+(k_1\lra k_2),
\nn\\[.5em]
\lefteqn{\de_3A^{--}_{+-}\Big|_{\mathrm{quad}} = 
- P_{\PW}(p_1+p_2)P_{\PW}(p_3+p_4) P_{\PW}(p_1+p_2-k_1)\cspbf\spac} &&
\nn\\
&& {}
\times \Biggl\{-\dkag^2\biggl[\cspbd\spae
+\frac{1}{2\MW^2}\cspcd\cspbf\spac\spef\biggr]
\nn\\
&& \qquad {}
+\dkag\frac{\lag}{\MW^2}\biggl[-2(p_1+p_2-k_1)^2\cspbd\spae
\nn\\
&& \qquad\qquad {}
+\langle p_2[K_2-K_1]p_3\rangle\langle k_2[P_1+P_2]k_1\rangle
-\cspcd\cspbf\spef\spac\biggr]
\nn\\
&& \qquad {}
+\frac{\lag^2}{\MW^4}\cspcd\spef\biggl[-\frac{1}{2}(p_1+p_2-k_1)^2 \cspbf\spac
\nn\\
&& \qquad\qquad {}
+\langle p_4[K_2-P_3]k_1\rangle
\langle k_2[K_1-P_2]p_1\rangle\biggr]\Biggr\}
\nn\\
&& 
+\Big(\{p_1,Q_1;p_2,Q_2\}\lra\{p_3,Q_3;p_4,Q_4\}\Big),
\nn\\[.5em]
\lefteqn{\de_3A^{--}_{++}\Big|_{\mathrm{lin}} = 
2 P_{\PW}(p_1+p_2)P_{\PW}(p_3+p_4)\spbc\spbe} &&
\nn\\
&& {}
\times \frac{\cspdf}{\cspad\cspaf}
\biggl[\dkag\cspdf-\frac{\lag}{\MW^2}\cspef\cspcd\spce\biggr]
\nn\\
&& {}
+\Biggl\{2(Q_4-Q_3) \left[-Q_1+(Q_1-Q_2)2(k_1p_1)P_{\PW}(p_1+p_2)\right]
P_{\PW}(p_3+p_4)
\nn\\
&&\qquad {} \times 
P_{\PW}(p_3+p_4-k_2)
\frac{ \spbe\langle p_2[P_1-K_1]k_2\rangle }{\cspac\cspad}
\nn\\
&&\qquad {}
\times\biggl[\dkag\cspdf+\frac{\lag}{\MW^2}\cspef
\langle p_2[P_4-K_2]p_3\rangle\biggr]\nn\\
&&\qquad {}
+\Big(\{p_1,Q_1;p_2,Q_2\}\lra\{p_3,Q_3;p_4,Q_4\}\Big)\Biggr\}
+(k_1\lra k_2),
\nn\\[.5em]
\lefteqn{\de_3A^{--}_{+-}\Big|_{\mathrm{lin}} = 
-2P_{\PW}(p_1+p_2)P_{\PW}(p_3+p_4)\spce
\frac{\spac\spae}{\spbc\spbe}} &&
\nn\\
&&{} \times
\biggl[-\dkag\cspdf+\frac{\lag}{\MW^2}
\Bigl(\cspcd\cspbf\spbc-\cspef\cspbd\spbe
\nn\\
&& \qquad {}
+\cspcd\cspef\spce\Bigr)\biggr]
\nn\\
&&{}
-\Biggl\{2(Q_1-Q_2) P_{\PW}(p_1+p_2)
\biggl[ \left[Q_4+(Q_3-Q_4)2(k_2p_4)P_{\PW}(p_3+p_4)\right]
\nn\\
&& \qquad {}
\times
P_{\PW}(p_1+p_2-k_1)
\frac{\spac\spae}{\spbe\spbf}
\Bigl( -\dkag\langle p_2[P_4-K_2]p_3\rangle
\nn\\
&& \qquad\qquad {}
+\frac{\lag}{\MW^2}\cspcd\spce(p_3+p_4-k_2)^2\Bigr)\biggr]
\nn\\
&& \qquad {}
+\Big(\{p_1,Q_1;p_2,Q_2\}\lra\{p_3,Q_3;p_4,Q_4\}\Big)\Biggr\}
\nn\\*
&& {}
+\Big(\mbox{c.c. and }\{p_1,Q_1;p_3,Q_3;k_1\}\lra \{p_2,Q_2;p_4,Q_4;k_2\}\Big),
\eeqar
where ``c.c.\ and $\{\dots\}\lra\{\dots\}$'' indicates that the complex conjugate
of the preceding expression has to be added after some substitutions.
The auxiliary functions for the remaining polarizations
follow from the relations \refeq{eq:CP} and \refeq{eq:P}.

In order to check our results, 
we have implemented the ATGC of the effective Lagrangian
\refeq{eq:Ltriple} into the program {\Madgraph} \cite{Stelzer:1994ta}
and compared our amplitudes
with the {\Madgraph} results for various phase-space points.
We found perfect numerical agreement.

\subsection{Amplitudes with genuine quartic gauge-boson couplings}
\label{se:qgc}

Figure~\ref{fig:AQGC} shows the only diagram with an AQGC 
(generically denoted by $a_4$) that contributes to $\ggffff$.
\bfi
\centerline{
\begin{picture}(120,100)(0,0)
\Photon(12,70)(45,45){2}{4}
\Photon(12,20)(45,45){2}{4}
\GCirc(45,45){5}{0}
\Photon(45,45)(80,70){2}{4}
\Photon(45,45)(80,20){2}{4}
\Vertex(80,70){2}
\ArrowLine(80,70)(100,80)
\ArrowLine(100,60)(80,70)
\Vertex(80,20){2}
\ArrowLine(80,20)(100,30)
\ArrowLine(100,10)(80,20)
\put(-5,70){$\ga_1$}
\put(-5,20){$\ga_2$}
\put(38,70){\small$\{\PW,\PZ\}$}
\put(38,15){\small$\{\PW,\PZ\}$}
\put(105,78){$f_1$}
\put(105,58){$\bar f_2$}
\put(105,28){$f_3$}
\put(105,8){$\bar f_4$}
\end{picture}}
\caption{Diagram with AQGC (black blob) contributing to $\ggffff$.}
\label{fig:AQGC}
\efi
For CC processes the ``anomalous diagram'' contributes in 
${\cal O}(a_4)$ to the cross section, because it is (as the SM
contribution) doubly resonant.
For NC processes, the diagram involves one power of $a_4$ and
two Z-boson resonances and interferes with the singly-resonant
SM amplitude. In this case, the contribution to the corresponding
cross section is suppressed by $a_4\GZ/\MZ$ w.r.t.\ CC
cross sections, i.e., the suppression factor involves one factor
in the anomalous coupling or in $\GZ/\MZ$ less than we counted
for NC ATGC. In the following we take both CC and NC AQGC
into account.

The AQGC contributions to the amplitudes read
\beqar
{\cal M}^{\si_1\si_2\si_3\si_4}_{\la_1\la_2,\ga\ga VV}
&=& \frac{e^4}{8\Lambda^2}
\de_{\si_1,-\si_2} \de_{\si_3,-\si_4}\,
g_{\ga\ga VV} \,
g^{\si_1}_{V\bar f_1 f_2} g^{\si_3}_{V\bar f_3 f_4}
P_{V}(p_1+p_2)P_{V}(p_3+p_4)
\nn\\
&& {} \times
\de_4 A_{\la_1\la_2}^{\si_1\si_3}(k_1,k_2,p_1,p_2,p_3,p_4)
\eeqar
with
\beq
g_{\ga\ga\PW\PW} = 1, \qquad
g_{\ga\ga\PZ\PZ} = \frac{1}{\cw^2}
\label{eq:gAAVV}
\eeq
and
\beqar
\label{eq:aqgc}
\de_4 A_{++}^{--}(k_1,k_2,p_1,p_2,p_3,p_4) &=& 
(4a_0-4\ri\tilde a_0+a_{\mathrm{c}})
\cspdf\spab^2\spce,
\nn \\
\de_4 A_{+-}^{--}(k_1,k_2,p_1,p_2,p_3,p_4) &=& 
-2a_{\mathrm{c}}\cspbd\cspbf\spac\spae.
\eeqar
The remaining auxiliary functions $\de_4 A_{\la_1\la_2}^{\si_1\si_3}$
can be obtained via the substitutions
\beqar
\de_4 A_{\la_1\la_2}^{\si_1,+}(k_1,k_2,p_1,p_2,p_3,p_4)
&=&
\de_4 A_{\la_1\la_2}^{\si_1,-}(k_1,k_2,p_1,p_2,p_4,p_3),
\nn \\
\de_4 A_{\la_1\la_2}^{+,\si_3}(k_1,k_2,p_1,p_2,p_3,p_4)
&=&
\de_4 A_{\la_1\la_2}^{-,\si_3}(k_1,k_2,p_2,p_1,p_3,p_4),
\nn \\
\de_4 A_{\la_1\la_2}^{\si_1\si_3}(k_1,k_2,p_1,p_2,p_3,p_4)
&=&
\left(\de_4 A_{-\la_1,-\la_2}^{-\si_1,-\si_3}
(k_1,k_2,p_1,p_2,p_3,p_4)\right)^{\ast}.
\label{eq:A4subst}
\eeqar
The generic amplitude $\M_{\la_1\la_2,\ga\ga VV}^{\si_1\si_2\si_3\si_4}$
is coherently added to the SM amplitude 
$\M_{\la_1\la_2,V}^{\si_1\si_2\si_3\si_4}$ of
Eq.~\refeq{genfun4f} for $V=\PW,\PZ$, respectively.
The colour summation of the squared amplitudes for
the various process types proceeds as in the SM case.

Again we have checked the amplitudes against results obtained with
{\Madgraph}, as explained at the end of the previous section. 

\section{Effective $\ga\ga\PH$ coupling and Higgs resonance}
\label{se:higgs}

In order to incorporate a possible Higgs resonance in 
$\gamma\gamma\to\PH\to\PV\PV\to 4f$ with $\PV=\PW,\PZ$, 
as depicted in \reffi{fig:higgs}, we consider 
an effective coupling of the Higgs boson to two photons.
\bfi
\centerline{
\begin{picture}(120,110)(0,0)
\Photon(12,70)(45,45){2}{4}
\Photon(12,20)(45,45){2}{4}
\GCirc(45,45){5}{0}
\DashLine(45,45)(85,45){5}
\Vertex(85,45){2}
\Photon(85,45)(120,70){2}{4}
\Photon(85,45)(120,20){2}{4}
\Vertex(120,70){2}
\ArrowLine(120,70)(140,80)
\ArrowLine(140,60)(120,70)
\Vertex(120,20){2}
\ArrowLine(120,20)(140,30)
\ArrowLine(140,10)(120,20)
\put(-5,70){$\ga_1$}
\put(-5,20){$\ga_2$}
\put(63,30){$\PH$}
\put(75,71){\small$\{\PW,\PZ\}$}
\put(75,13){\small$\{\PW,\PZ\}$}
\put(145,78){$f_1$}
\put(145,58){$\bar f_2$}
\put(145,28){$f_3$}
\put(145,8){$\bar f_4$}
\end{picture}}
\caption{Diagram with effective $\ga\ga\PH$ coupling (black blob).}
\label{fig:higgs}
\efi
In the SM this coupling is mediated via fermion (mainly top-quark)
and W-boson loops.
We define the effective Lagrangian for the $\ga\ga\PH$ vertex 
\cite{Shifman:eb} by
\beq
\L_{\ga\ga\PH}=-\frac{g_{\ga\ga\PH}}{4} F^{\mu\nu}F_{\mu\nu}\frac{H}{v},
\eeq
where $v=2\MW\sw/e$ is the vacuum expectation value of the Higgs field $H$.
Up to normalization, $\L_{\ga\ga\PH}$ is the lowest-dimensional,
CP-conserving, electromagnetically gauge-invariant operator for two
photons and the scalar field $H$.
The corresponding Feynman rule reads
\beq
\ri\Ga^{\ga\ga\PH}_{\mu\nu}(k_1,k_2,k_\PH) =
\frac{\ri g_{\ga\ga\PH}}{v}
\left[g_{\mu\nu}(k_1k_2)-k_{1,\nu}k_{2,\mu}\right],
\eeq
where $k_1,k_2$ are the incoming photon momenta.
Comparing this Feynman rule to the loop-induced SM vertex with
the external fields on shell, which 
has, e.g., been given in \citeres{Denner:1995jv,Shifman:eb}, 
we obtain 
\beqar
\label{eq:higgs}
g_{\ga\ga\PH}\Big|_{\SM} &=& \frac{\al}{\pi}
\Biggl\{\frac{6\MW^2}{\MH^2}+1
+\frac{6\MW^2}{\MH^2}(2\MW^2-\MH^2)
C_0(\MH,\MW)
\nn\\
&& \qquad {}
-2\sum_f{N_f^{\mathrm{c}} Q_f^2 \frac{m_f^2}{\MH^2}\left[2+(4m_f^2-\MH^2)
C_0(\MH,m_f)\right]}\Biggr\},
\eeqar
where the colour factor $N_f^{\mathrm{c}}$ in the sum over all fermions
$f$ is equal to 3 for quarks and 1 for leptons.
The scalar 3-point integral $C_0$ is given by
\beq
C_0(\MH,m)=\frac{1}{2\MH^2}\ln^2\left(\frac{\be_m+1}{\be_m-1}\right),
\qquad
\be_m=\sqrt{1-\frac{4m^2}{\MH^2}+\ri 0}.
\eeq
The complete matrix elements for the 
diagrams with a Higgs resonance (as shown in \reffi{fig:higgs}) 
can then be written as 
\beqar
{\cal M}^{\si_1\si_2\si_3\si_4}_{\la_1\la_2,\PH VV}
&=& -\frac{e^4}{2\sw^2} \, \de_{\si_1,-\si_2} \de_{\si_3,-\si_4}\,
g_{\ga\ga\PH} \, g_{\ga\ga VV} \,
g^{\si_1}_{V\bar f_1 f_2} g^{\si_3}_{V\bar f_3 f_4}
P_{V}(p_1+p_2)P_{V}(p_3+p_4)
\nn\\
&& {} \times
\MH^2 P_\PH(k_1+k_2) \,
\de_\PH A_{\la_1\la_2}^{\si_1\si_3}(k_1,k_2,p_1,p_2,p_3,p_4)
\eeqar
with $g_{\ga\ga VV}$ defined in Eq.~\refeq{eq:gAAVV} and
\beqar
\de_\PH A_{++}^{--}(k_1,k_2,p_1,p_2,p_3,p_4) &=& 
\frac{\spab}{\cspab}\cspdf\spce,
\qquad
\de_\PH A_{\pm\mp}^{\si_1\si_3} = 0.
\eeqar
The other expressions for $\de_\PH A_{\la_1\la_2}^{\si_1\si_3}$
follow in the same way as described in Eq.~\refeq{eq:A4subst}
for $\de_4 A_{\la_1\la_2}^{\si_1\si_3}$.
The width in the Higgs-boson propagator $P_{\PH}$ is introduced 
in the same way as in \refse{se:finwidth} for the gauge bosons.

\section{Monte Carlo integration}
\label{sec:integration}

The squared matrix element is integrated over the phase space 
following the strategy described in 
\citeres{Denner:1999gp,Dittmaier:2002ap,Roth:kk}. 
We apply the multi-channel Monte Carlo technique
\cite{Berends:1994pv} where the integrand is flattened 
by choosing appropriate mappings of the pseudo-random numbers
into the momenta of the outgoing particles. 
This is necessary because the 
integrand shows a very complex peaking structure in eight 
($\ggffff$) or eleven ($\ggffffg$) dimensions. 
This complicated structure of the
integrand (induced by various diagram types) deteriorates the 
statistical error of the numerical integration and can lead to 
numerically unstable results. 
For each diagram we introduce an own 
appropriate mapping, so-called ``channels'', so that
more events are generated in regions where the squared matrix element 
of the diagram is large. 
The multi-channel approach combines these channels
in such a way that the whole integrand is widely smoothed everywhere.
For each event a single channel is randomly chosen, and the phase-space 
configuration is determined according to the mapping of this channel.
The probability to choose a given channel, called a-priori weight,
is optimized according to \citere{Kleiss:qy} to minimize 
the statistical error as much as possible.

The convolution over the photon spectrum is given by
\beq
\rd\si=\int_0^1\rd x_1\int_0^1 \rd x_2\, f_\gamma(x_1)\,f_\gamma(x_2)\,
\rd\si_{\ga\ga}(x_1P_1,x_2P_2),
\eeq
where $\rd\si_{\ga\ga}$ is the differential $\ga\ga$ cross section.
The function 
$f_\gamma(x_i)$ denotes the probability density for getting a photon with
momentum $k_i=x_i P_i$, and $P_i$ is the electron momentum before
Compton backscattering.
In order to reduce the statistical error of this integration we use 
a simple way of stratified sampling. 
The integration region for $x_i$ of each photon spectrum is divided into 
a fixed number of bins. We choose bin $i$ with a probability 
$\al_i$ and divide the corresponding weight by $\al_i$. In this way 
the integration remains formally unchanged if we normalize 
$\sum_i \alpha_i=1$. The parameters $\al_i$ can be 
used to improve the convergence of the numerical integration.
By choosing the $\al_i$ proportional to the cross section 
of the corresponding bin $i$,
more events are sampled in regions where the photon spectrum is large. 
Care has to be taken that the $\al_i$ do not become too small because 
this might lead to rare events with very large weights that render the 
error estimate unreliable. This optimization typically reduces the 
Monte Carlo integration error by a factor 2--5.

\section{Numerical results}
\label{se:numerics}
\subsection{Input parameters}

We use the following set of input parameters \cite{Hagiwara:fs}:
\beqar
\begin{array}[b]{r@{\,}lr@{\,}l}
\MW &= 80.423\GeV, \qquad & \GW &= 2.118\GeV, \\
\MZ &= 91.1876\GeV, & \GZ &= 2.4952\GeV, \\
M_{\PH} &= 170\GeV, & \GH &= 0.3834\GeV, \\
\al(0) &= 1/137.03599976, & \alpha_{\mathrm{s}} &= 1.1172,\\
G_{\mu} &= 1.16639\times 10^{-5}\GeV^{-2},
\end{array}
\eeqar
where the Higgs mass is chosen well above the W-pair threshold so that
intermediate Higgs bosons decay rapidly into W pairs; the corresponding 
decay width $\GH$ has been obtained with the program 
{\HDECAY} \cite{Djouadi:1997yw}.

Furthermore, we apply the separation cuts
\beqar
\label{eq:cuts}
\begin{array}[b]{r@{\,}lr@{\,}lr@{\,}lr@{\,}l}
E_{\gamma}&>10\GeV, \qquad & \theta(\gamma,\mathrm{beam})&>5^{\circ}, \qquad &
\theta(l,\gamma)&>5^{\circ}, \qquad & \theta(q,\gamma)&>5^{\circ}, \\
E_l&>10\GeV, & \theta(l,\mathrm{beam})&>5^{\circ}, &
\theta(l,l')&>5^{\circ}, & \theta(l,q)&>5^{\circ}, \\
E_q&>10\GeV, & \theta(q,\mathrm{beam})&>5^{\circ}, & 
m(q,q')&>10\GeV,
\end{array}
\eeqar
where $q$ and $l$ denote quarks and charged leptons, respectively, 
and $m(q,q')$ is the invariant mass of an outgoing quark pair.
The energies $E_X$ and angles $\theta(X,Y)$ are defined in the laboratory 
frame. 
Using these cuts all infrared, i.e., soft or collinear, singularities 
are removed from the phase space. 

In order to account for leading universal corrections,
we use two different values for the coupling constant 
$\alpha=e^2/(4\pi)$. Since on-shell photons couple to 
charged particles with the coupling constant $\alpha(0)$
(effective electromagnetic coupling at zero-momentum transfer),
we take this coupling for each external photon in the processes $\ggffff$ 
and $\ggffffg$. 
For CC reactions, the remaining 
couplings correspond to $\PW f \bar{f}$ vertices. 
For these vertices a large part of the electroweak radiative corrections
\cite{Bardin:1986fi}
(the running of the electromagnetic coupling and
the universal corrections related to the $\rho$ parameter)
are absorbed into an effective electromagnetic coupling $\alpha_{G_{\mu}}$ 
which is derived from the Fermi constant $\GF$ by
\beq
\alpha_{G_{\mu}}=\frac{\sqrt{2}G_{\mu}\MW^2\sw^2}{\pi}.
\eeq
Therefore, in the following numerical studies, 
we replace $\alpha^4$ by 
$\alpha(0)^2\alpha_{G_{\mu}}^2$ for the processes $\ggffff$ and 
$\alpha^5$ by $\alpha(0)^3\alpha_{G_{\mu}}^2$ for $\ggffffg$.

For the evaluation of the photon spectrum we use the program 
{\CompAZ} \cite{Zarnecki:2002qr} with the polarization of the laser beams 
$-1$ (i.e.\ photon helicity $-1$)%
\footnote{Internally in {\CompAZ} the polarization of the laser light is 
defined as the negative of the photon helicity.}
and the polarization of the electron beams $+0.85$. 
This choice for the relative signs in the polarizations yields a
sharper peak at the upper end of the photon spectrum.
Results for monochromatic photon beams are always shown for
unpolarized photons.

If not stated otherwise, the results are obtained in the fixed-width scheme.

The numerical integration is carried out using $10^7$ events. 
The runtime of our Monte Carlo program on a PC with $2\,\mathrm{GHz}$ varies 
from 30 minutes to 6 hours depending on the considered process.

\subsection{Results for integrated cross sections}
\subsubsection{Survey of cross sections}
\label{se:cross}

\renewcommand{\arraystretch}{1.3}
\begin{table}
\centerline{
\begin{tabular}{|c||c|c||c|c|}
\hline
&\multicolumn{2}{c||}{present work} & \multicolumn{2}{c|}{{\Whizard}/{\Madgraph}}
\\ \cline{2-5}
$\ga\ga\to$ & $\si_{4f}[\fb]$ & $\si_{4f}[\fb](\mathrm{conv})$ & 
$\si_{4f}[\fb]$ & $\si_{4f}[\fb](\mathrm{conv})$
\\ \hline 
$\Pem\Pnebar\Pnmu\Pmup$& 826.47(21)  & 190.87(10) & 826.39(26) & 191.05(16)
\\ \hline
$\Pem\Pep\Pnmu\Pnmubar$& 1.75460(62) & 0.90525(61) & 1.75518(78) & 0.9050(11)
\\ \hline
$\Pem\Pep\Pmum\Pmup$   & 19.400(33)  & 19.129(61) & 19.342(21) & 19.188(48)
\\ \hline
$\Pem\Pep\Pem\Pep$     & 9.469(17)   & 9.357(32)  & 9.453(11) & 9.383(25)
\\ \hline
$\Pem\Pep\Pne\Pnebar$  & 828.34(21)  & 191.72(10) & 828.29(26) & 191.55(17)
\\ \hline
$\Pem\Pnebar\Pu\Pdbar$ & 2351.11(68) & 565.05(33) & 2351.79(84) & 565.07(51)
\\ \hline
$\Pne\Pep\Pd\Pubar$    & 2350.84(68) & 558.39(32) & 2353.21(84) & 558.41(50)
\\ \hline
$\Pne\Pnebar\Pu\Pubar$ & 1.19761(50) & 0.61256(50) & 1.19684(57) & 0.61083(71)
\\ \hline
$\Pne\Pnebar\Pd\Pdbar$ & 0.095981(44)& 0.049092(45) & 0.096011(48) & 0.049118(57)
\\ \hline
$\Pem\Pep\Pu\Pubar$    & 14.036(21)  & 10.597(26) & 14.016(15) & 10.574(21)
\\ \hline
$\Pem\Pep\Pd\Pdbar$    & 4.7406(29)  & 2.6614(32) & 4.7377(28) & 2.6651(38)
\\ \hline
$\Pu\Pdbar\Ps\Pcbar$   & 6659.6(2.1) & 1603.8(1.0) & 6663.5(2.7) & 1605.0(1.5)
\\ \hline
$\Pu\Pubar\Pc\Pcbar$   & 10.469(14)  & 6.111(12) & 10.4531(88) & 6.113(10)
\\ \hline
with QCD               & 1543.6(2.9) & 1071.3(2.9) & --- & ---
\\ \hline
$\Pu\Pubar\Ps\Psbar$   & 3.3282(21)  & 1.6569(18) & 3.3310(20) & 1.6595(23)
\\ \hline
with QCD               & 412.97(75)  & 288.79(72) & --- &  ---
\\ \hline
$\Pd\Pdbar\Ps\Psbar$   & 0.49807(29) & 0.23232(24) & 0.49804(30) & 0.23252(32)
\\ \hline
with QCD               & 96.34(18)   & 66.80(18) &---  &  ---
\\ \hline
$\Pu\Pubar\Pu\Pubar$   & 5.1846(69)  & 3.0298(57) & 5.1900(45) & 3.0419(53)
\\ \hline
with QCD               & 772.6(1.5)  & 538.9(1.4) & --- &  ---
\\ \hline
$\Pd\Pdbar\Pd\Pdbar$   & 0.24683(15) & 0.11581(12) & 0.24665(17) & 0.11579(17)
\\ \hline
with QCD               & 48.252(96)  & 33.685(88) & --- &  ---
\\ \hline
$\Pu\Pubar\Pd\Pdbar$   & 6663.5(2.3) & 1606.1(1.1) & 6664.8(2.8) & 1604.6(1.6)
\\ \hline
with QCD               & 7075.8(3.7) & 1896.4(2.9) & --- &  ---
\\ \hline
\end{tabular} }
\caption{Total cross sections for $\gamma\gamma\rightarrow 4f$  
at $\sqrt{s}=500 \GeV$ for various final states with and without 
convolution over the photon spectrum.}
\label{tab:cross4f}
\end{table}

\renewcommand{\arraystretch}{1.3}
\begin{table}
\centerline{
\begin{tabular}{|c||c|c||c|c|}
\hline
&\multicolumn{2}{c||}{present work} & \multicolumn{2}{c|}{{\Whizard}/{\Madgraph}}
\\ \cline{2-5}
$\ga\ga\to$ & $\si_{4f\ga}[\fb]$ & $\si_{4f\ga}[\fb](\mathrm{conv})$ & 
$\si_{4f\ga}[\fb]$ & $\si_{4f\ga}[\fb](\mathrm{conv})$
\\ \hline 
$\Pem\Pnebar\Pnmu\Pmup\ga$& 39.234(44)    & 6.188(11)     & 39.218(29) & 6.2040(87)
\\ \hline
$\Pem\Pep\Pnmu\Pnmubar\ga$& 0.10157(10)   & 0.028612(40)  & 0.101556(88)& 0.028548(52)
\\ \hline
$\Pem\Pep\Pmum\Pmup\ga$   & 1.0567(35)    & 0.5083(28)    & 1.0547(20) & 0.5091(29)
\\ \hline
$\Pem\Pep\Pem\Pep\ga$     & 0.5085(18)    & 0.2433(13)    & 0.5091(10) & 0.2461(12)
\\ \hline
$\Pem\Pep\Pne\Pnebar\ga$  & 39.301(46)    & 6.213(11)     & 39.332(30) & 6.2069(89)
\\ \hline
$\Pem\Pnebar\Pu\Pdbar\ga$ & 96.61(13)     & 14.216(27)    & 96.575(75) & 14.159(21)
\\ \hline
$\Pne\Pep\Pd\Pubar\ga$    & 96.60(13)     & 15.459(30)    & 96.520(76) & 15.429(22)
\\ \hline
$\Pne\Pnebar\Pu\Pubar\ga$ & 0.030818(35)  & 0.008640(14)  & 0.030756(28) & 0.008609(16)
\\ \hline
$\Pne\Pnebar\Pd\Pdbar\ga$ & 0.00061753(75)& 0.00017313(31)& 0.00061731(56) & 0.00017358(34)
\\ \hline
$\Pem\Pep\Pu\Pubar\ga$    & 0.6446(17)    & 0.25463(99)   & 0.6477(10) & 0.2579(10)
\\ \hline
$\Pem\Pep\Pd\Pdbar\ga$    & 0.26653(36)   & 0.08137(17)   & 0.26689(28) & 0.08166(21)
\\ \hline
$\Pu\Pdbar\Ps\Pcbar\ga$   & 229.86(36)    & 32.621(81)    & 229.52(19) & 32.531(49)
\\ \hline
$\Pu\Pubar\Pc\Pcbar\ga$   & 0.30556(69)   & 0.10718(34)   & 0.30563(47) & 0.10836(43)
\\ \hline
with QCD                  & 34.73(14)     & 13.801(77)    & --- & ---
\\ \hline
$\Pu\Pubar\Ps\Psbar\ga$   & 0.08791(13)   & 0.026278(59)  & 0.087935(98) & 0.026271(65)
\\ \hline
with QCD                  & 6.362(23)     & 2.493(13)     &---  & ---
\\ \hline
$\Pd\Pdbar\Ps\Psbar\ga$   & 0.0046253(71) & 0.0014842(37) & 0.0046191(52) & 0.0014832(36)
\\ \hline
with QCD                  & 0.5427(22)    & 0.2165(11)    & --- & ---
\\ \hline
$\Pu\Pubar\Pu\Pubar\ga$   & 0.15081(33)   & 0.05301(16)   & 0.15082(21) & 0.05332(16)
\\ \hline
with QCD                  & 17.377(71)    & 6.964(35)     & --- & ---
\\ \hline
$\Pd\Pdbar\Pd\Pdbar\ga$   & 0.0022893(37) & 0.0007421(21) & 0.0022878(25) & 0.0007398(18)
\\ \hline
with QCD                  & 0.2716(11)    & 0.10863(53)   & --- & ---
\\ \hline
$\Pu\Pubar\Pd\Pdbar\ga$   & 229.86(40)    & 32.85(15)     & 229.65(19) & 32.518(51)
\\ \hline
with QCD                  & 236.31(42)    & 35.14(11)     & --- & ---
\\ \hline
\end{tabular} }
\caption{Total cross sections for $\gamma\gamma\rightarrow 4f\ga$  
at $\sqrt{s}=500 \GeV$ for various final states with and without
convolution over the photon spectrum.}
\label{tab:cross4fa}
\end{table}

In order to illustrate the reliability of our Monte Carlo generator we 
compare our results on cross sections for a representative set of 
the processes $\ggffff$ and $\ggffffg$ with the results obtained with 
the Monte Carlo program {\Whizard} (version 1.28) \cite{Kilian:2001qz} 
which uses the matrix-element generator {\Madgraph} \cite{Stelzer:1994ta}%
\footnote{For a tuned comparison we rescaled the {\Whizard}/{\Madgraph} 
results by a factor $\alpha(0)^2\alpha_{G_{\mu}}^2/\alpha^4$ for $\ggffff$ 
and $\alpha(0)^3\alpha_{G_{\mu}}^2/\alpha^5$ for $\ggffffg$.}.
In \reftas{tab:cross4f} and \ref{tab:cross4fa} we list the results 
for the 17 different final states defined in \refta{ta:classes}. 
The numbers in parentheses correspond to the Monte Carlo error.
For the final states that can be produced via 
intermediate gluons we compute the cross section both with and 
without gluon-exchange contributions. Since the version of {\Madgraph} 
implemented in {\Whizard} is not able to deal with interferences of 
electroweak and QCD diagrams, we give only the pure 
electroweak {\Whizard}/{\Madgraph} results for these processes. 
Furthermore, we list the corresponding cross sections 
with and without convolution over the photon beam spectrum. 
For this study, we have implemented the program {\CompAZ} into {\Whizard}.

As explained in \refse{se:pclass}, the cross sections for the
CP-equivalent final states
$\Pem\Pnebar\Pu\Pdbar(\ga)$ and $\Pne\Pep\Pd\Pubar(\ga)$
are not identical if the convolution over the photon beam spectrum is
carried out.
Therefore, we give results for both final states. 
In all other cases, the 
cross sections for a given final state and for the 
CP-conjugated one coincide.

CC and CC/NC processes possess the largest cross sections
because of the dominance of W-pair production.
The convolution over the photon spectrum reduces these cross sections
significantly since low-energy photons cannot 
produce on-shell W pairs. 
NC processes are affected less, and in 
some cases, such as $\ga\ga\to\Pep\Pem\Pmup\Pmum$, the cross section is only 
slightly reduced. 
Owing to the colour factors of the quarks, hadronic and semi-leptonic 
cross sections differ by roughly a factor $3$, hadronic and 
leptonic cross sections by roughly a factor $3^2=9$. 
For CC processes $\ggffff$ we obtain a rough estimate of the cross sections
by multiplying the cross section of $\ggtoww$ with
the branching ratios of the W~bosons into leptons or quarks
depending on the final state. Note that this estimate, 
which is only good within $10{-}20\%$, does not take
into account contributions from background diagrams, width effects, and 
cuts on final-state fermions.
The difference of cross sections for CC processes and the corresponding 
processes of mixed type reflects the size of the background contributions
induced by NC diagrams.

The results of {\Whizard}, which are also generated with $10^7$ events,
and of our program typically
agree within 1--2 standard errors.
The size of the statistical errors obtained with
{\Whizard} and our program is comparable.
The runtime of {\Whizard} is usually somewhat bigger than the one of 
our program. 
Depending on the process class, the speed of our program is $1{-}7$
times higher, where the largest difference occurs for NC processes.

\subsubsection{Energy dependence of integrated cross sections}
\label{se:energydep}

\begin{figure}
\setlength{\unitlength}{1cm}
\centerline{
\begin{picture}(7,8)
\put(-1.7,-14.5){\includegraphics{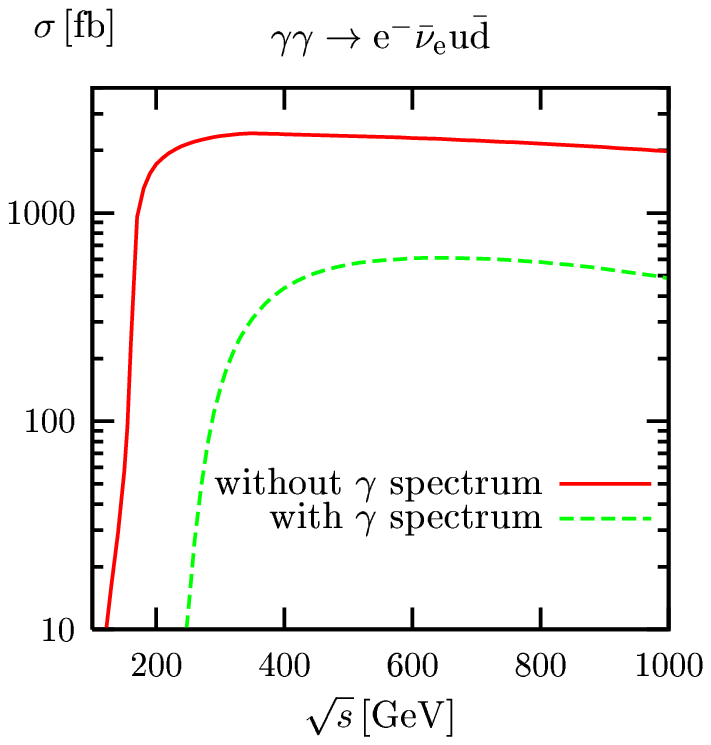}}
\end{picture}\hspace*{1em}
\begin{picture}(7,8)
\put(-1.7,-14.5){\includegraphics{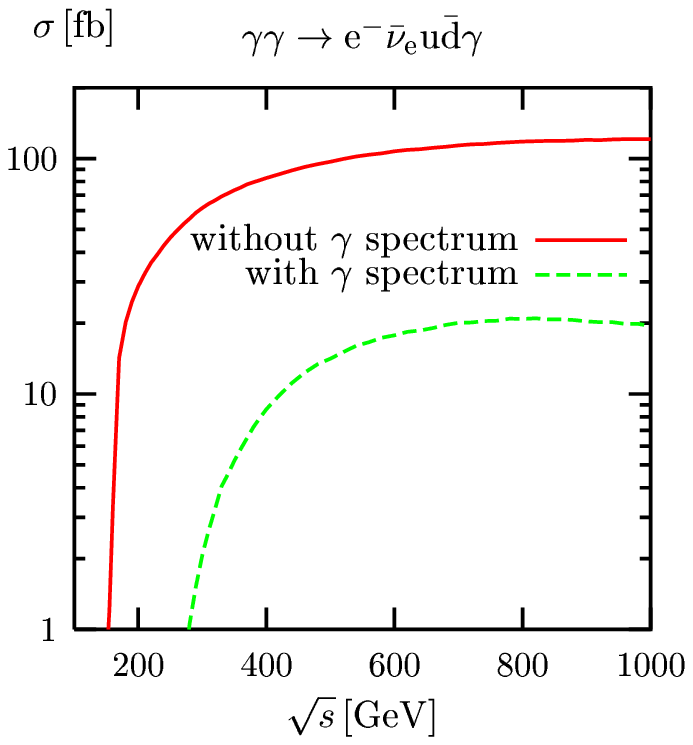}}
\end{picture}}
\caption{Integrated cross sections of the processes 
$\gamma\gamma\to\Pem\bar\Pne\Pu\bar\Pd(\ga)$ with and without 
convolution over the photon spectrum as a function of the CM energy 
$\sqrt{s}$.}
\label{fig:cme_enueud}
\end{figure}

In \reffi{fig:cme_enueud} we show the cross sections for the processes 
$\gamma\gamma\to\Pem\bar\Pne\Pu\bar\Pd(\ga)$ as a function of the 
centre-of-mass (CM) energy $\sqrt{s}$  
with and without convolution over the photon spectrum. 
Here and in the following,
with convolution over the photon spectrum $\sqrt{s}$ stands for the
CM energy $\sqrt{s_{\Pe\Pe}}$ of the incoming electron beams,
without convolution it is the CM energy $\sqrt{s_{\ga\ga}}$ of the
incoming photons.
In the case without photon spectrum, the rise of the cross section is 
clearly visible at the W-pair threshold, 
$\sqrt{s_{\ga\ga}}\gsim 160\GeV$. 
For $\gamma\gamma\to\Pem\bar\Pne\Pu\bar\Pd$ the cross section
increases roughly proportional to $\be=\sqrt{1-4\MW^2/s_{\ga\ga}}$ above 
the threshold, as expected from the two-particle phase space of the W pairs.
For $\gamma\gamma\to\Pem\bar\Pne\Pu\bar\Pd\gamma$ the rise of the 
cross section is not as steep because of the higher-dimensional 
$\PW\PW\ga$ phase space. 
The convolution over the photon spectrum reduces the 
available energy for W-pair production and shifts the onset of the 
cross section to higher CM energies.

The cross sections for $\ggffff$ as well as $\ggffffg$ decrease at 
high energies, even though the total cross section of the 
$\ggtoww$ process approaches a constant in the high-energy limit 
if no cuts are imposed, i.e., if the W~bosons are 
allowed to go in the beam directions. 
At high energies, however, forward and backward scattering of W~bosons
is restricted due to the cuts applied to the outgoing fermions, because
the decay fermions mainly follow the direction of the decaying W~boson.

\subsubsection{Contributions from CC, NC, and gluon-exchange diagrams} 

\begin{figure}
\setlength{\unitlength}{1cm}
\centerline{
\begin{picture}(7,8)
\put(-1.7,-14.5){\includegraphics{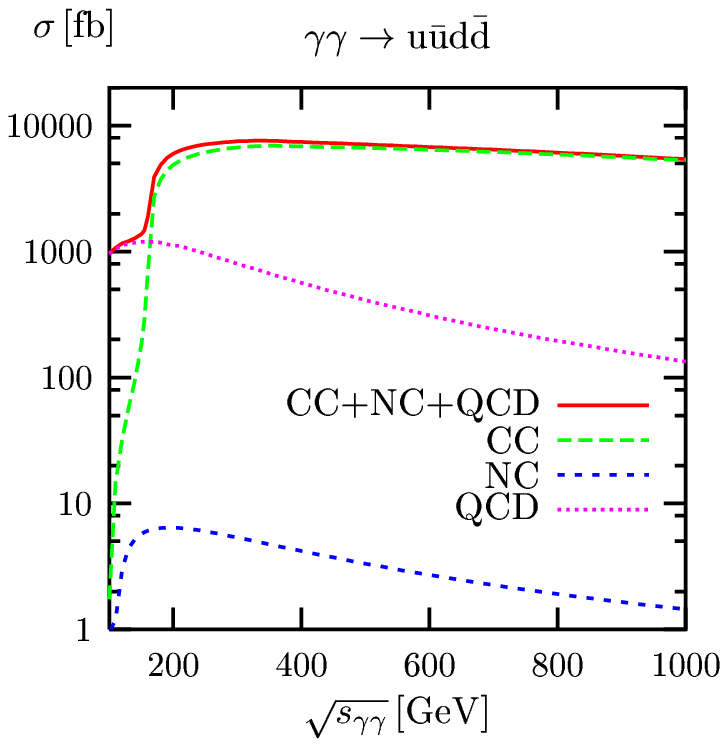}}
\end{picture}\hspace*{1em}
\begin{picture}(7,8)
\put(-1.7,-14.5){\includegraphics{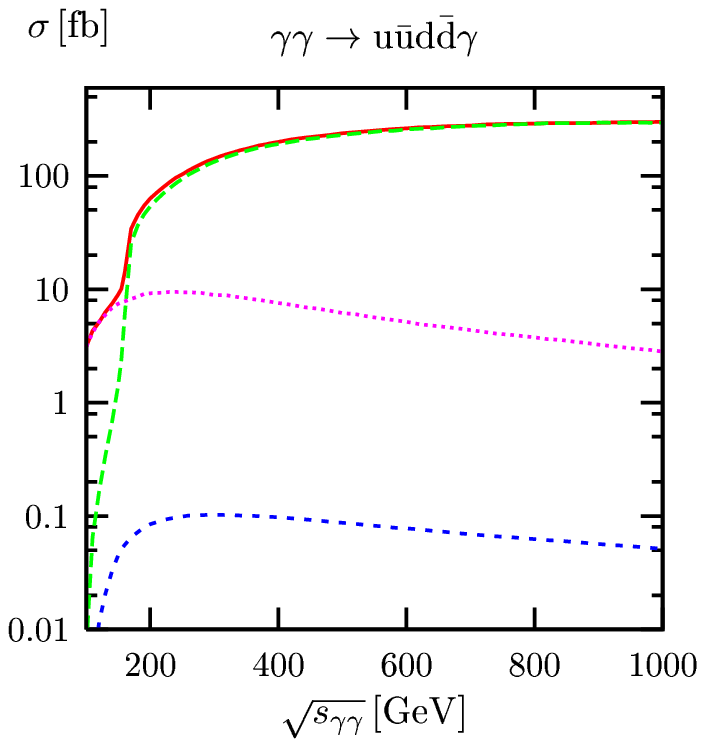}}
\end{picture}}
\caption{Different contributions to the integrated cross sections 
for the processes $\gamma\gamma\to\Pu\bar\Pu\Pd\bar\Pd(\ga)$ 
as a function of the CM energy without photon spectrum.}
\label{fig:cme_uudd}
\end{figure}

In \reffi{fig:cme_uudd} we show 
the impact of CC, NC, and gluon-exchange diagrams
on the CC/NC processes $\gamma\gamma\to\Pu\bar\Pu\Pd\bar\Pd$ and 
$\gamma\gamma\to\Pu\bar\Pu\Pd\bar\Pd\ga$.
We do not include the photon spectrum in this analysis.
Above the W-pair threshold, $\sqrt{s_{\ga\ga}}>160\GeV$, the CC diagrams 
are clearly dominating, while the contributions from gluon-exchange 
diagrams are one or two orders of magnitude smaller. 
The impact of the gluon-exchange diagrams strongly depends on 
the choice of the invariant-mass cut between two quarks, and
gluon-exchange diagrams are more important if the invariant-mass cut 
is small. The contributions from pure NC diagrams are totally negligible
as long as W-pair production is possible.

\subsubsection{W-pair signal diagrams and double-pole approximation}

\begin{figure}
\setlength{\unitlength}{1cm}
\centerline{
\begin{picture}(7,6.9)
\put(-1.2,-13.1){\includegraphics{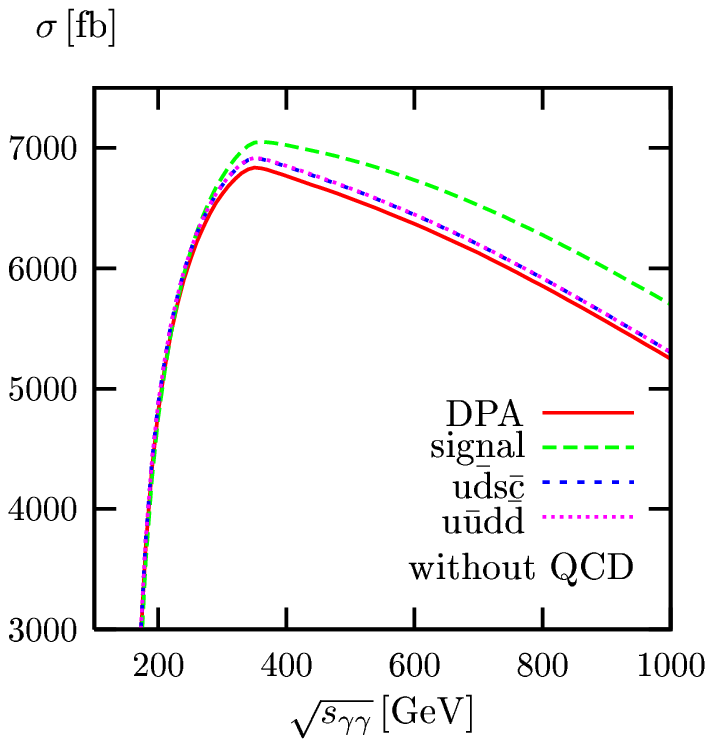}}
\end{picture}\hspace*{1em}
\begin{picture}(7,6.9)
\put(-1.2,-13.1){\includegraphics{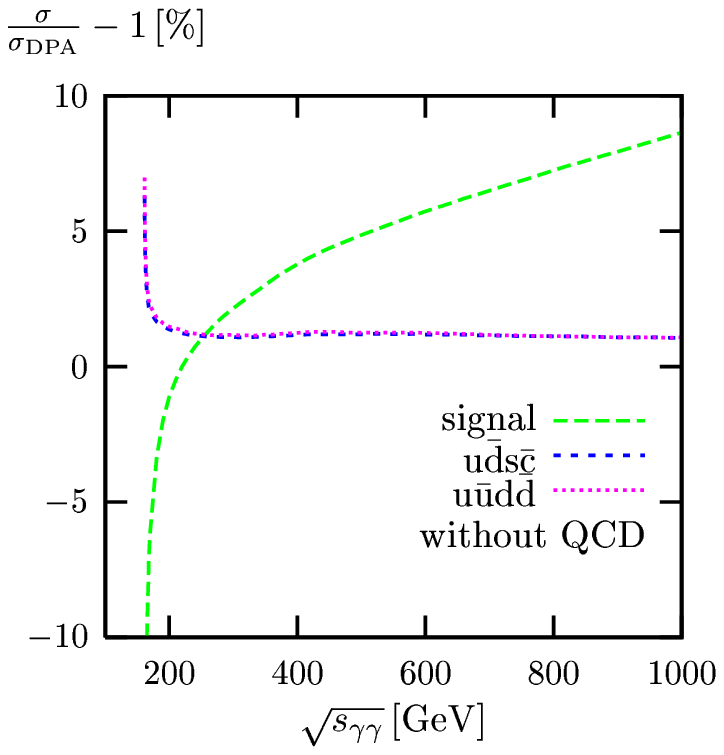}}
\end{picture}} 
\centerline{
\begin{picture}(7,6.9)
\put(-1.2,-13.1){\includegraphics{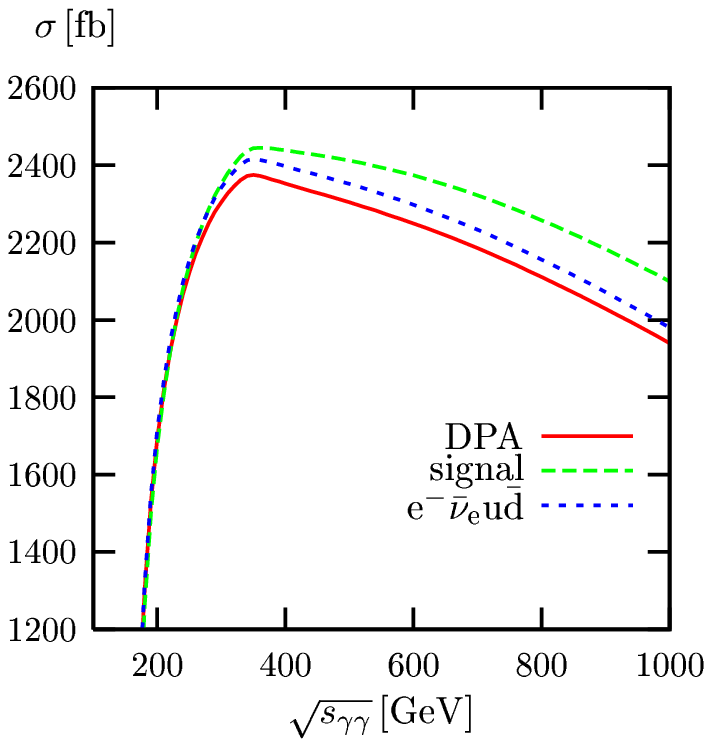}}
\end{picture}\hspace*{1em}
\begin{picture}(7,6.9)
\put(-1.2,-13.1){\includegraphics{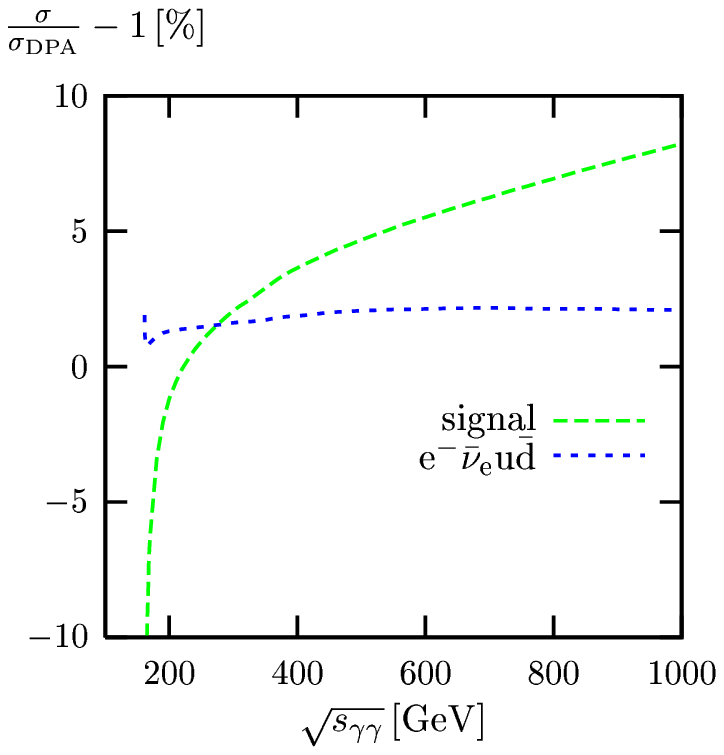}}
\end{picture}}
\centerline{
\begin{picture}(7,6.9)
\put(-1.2,-13.1){\includegraphics{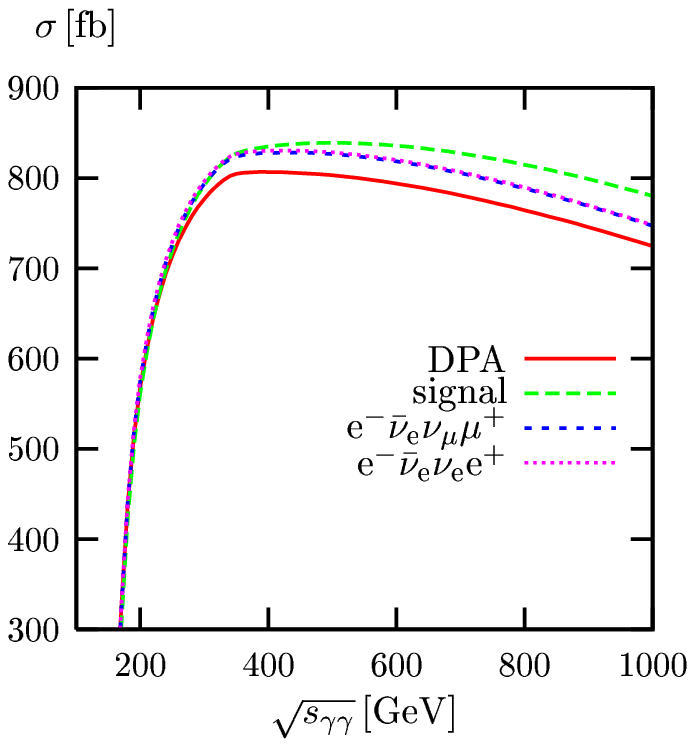}}
\end{picture}\hspace*{1em}
\begin{picture}(7,6.9)
\put(-1.2,-13.1){\includegraphics{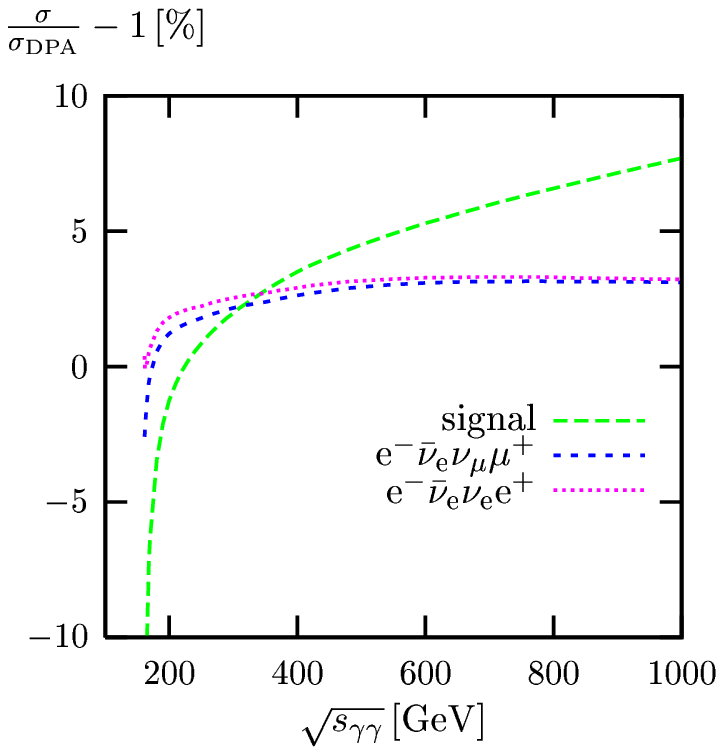}}
\end{picture}}
\caption{Cross sections of various processes including all diagrams, 
only W-pair 
signal diagrams, and in DPA as a function of the CM energy (l.h.s.), 
and the corresponding relative deviations from the DPA (r.h.s.); 
photon spectrum and gluon-exchange diagrams are not included.}
\label{fig:dpa}
\end{figure}

In \reffi{fig:dpa}
the cross sections of the W-pair signal diagrams and the 
DPA for $\ggwwffff$ (see \refse{se:dpa} 
for definitions) are compared with the complete lowest-order cross 
section for several processes. 
The plots on the l.h.s.\ show the cross sections for various final 
states calculated from the full set of (electroweak) diagrams,
from the signal diagrams only, and 
in DPA separately for hadronic, 
semi-leptonic, and leptonic final states, while the plots on the 
r.h.s.\ show the relative deviation from the corresponding DPA.
We do not include the convolution over the 
photon spectrum and gluon-exchange diagrams in this analysis 
so that effects of the approximation are clearly visible.
For energies not too close to the W-pair threshold,
the DPA agrees with the full lowest-order cross section within 
1--3\%, which is of the expected order of $\GW/\MW$.
Near threshold, i.e.\ for $\sqrt{s_{\ga\ga}}-2\MW={\cal O}(\GW)$,
the reliability of the DPA breaks down, since background
diagrams become more and more important and small scales $\ga$,
such as $\sqrt{s_{\ga\ga}-4\MW^2}$, can increase the naive error
estimate from $\GW/\MW$ to $\GW/\ga$.
The cross sections of the W-pair signal diagrams, however, shows large 
deviations from the full $\ggffff$ cross sections for the whole
energy range, in particular, at high 
energies. As explained in \refse{se:dpa}, the W-pair signal diagrams 
are not gauge invariant,
and thus the reliability and usefulness of the resulting predictions
should be investigated 
carefully. The results of \reffi{fig:dpa} clearly show that a naive
signal definition is a bad concept for $\ggwwffff$, since deviations
from the full process $\ggffff$ 
even reach 5--10\% in the TeV range. This is in contrast to the
situation at $\Pep\Pem$ colliders where the naive W-pair signal 
(defined in `t~Hooft--Feynman gauge) was a
reasonable approximation
(see, e.g., \citere{Grunewald:2000ju}).

The failure of the naive W-pair signal definition for
$\ga\ga$ collisions was also pointed out in 
\citeres{Moretti:1996nv,Baillargeon:1997se} before. 
In \citere{Baillargeon:1997se} an
``improved narrow-width approximation'' was presented which
provides another variant for a gauge-invariant W-pair signal definition.
It is based on the factorization of production and decay matrix elements,
while retaining W-spin correlations.

\subsubsection{Comparison of schemes for introducing finite gauge-boson widths}
\label{se:width}

In this section we compare the different implementations of
gauge-boson widths described in \refse{se:finwidth} numerically.
As explained in \refse{se:finwidth},
the complex-mass scheme is the only scheme that yields 
gauge-invariant results in general, but for the process classes
$\ggffff(\ga)$ the fixed-width approach (in the non-linear gauge)
also yields amplitudes that respect Ward identities and gauge cancellations.
Table~\ref{tab:width} lists the cross sections for 
the processes $\ga\ga\rightarrow\Pem\Pnebar\Pnmu\Pmup$ and 
$\ga\ga\rightarrow\Pem\Pnebar\Pnmu\Pmup\ga$ obtained 
with the fixed W~width,
the step-width, the running-width, and with the complex-mass scheme.
\begin{table}
\centerline{
\begin{tabular}{|c|c|c|c|c|c|}
\hline
\multicolumn{6}{|c|}{$\si(\ga\ga\rightarrow\Pem\Pnebar\Pnmu\Pmup)$}\\
\hline 
$\sqrt{s_{\ga\ga}}\,[\GeV]$ & 500 & 800 & 1000 & 2000 & 10000 \\
\hline
fixed width    & 826.40(21) & 788.35(21) & 746.94(21) & 500.70(20) & 31.745(68) \\
\hline
step width     & 827.45(22) & 789.34(21) & 748.17(23) & 501.41(21) & 31.746(68) \\
\hline
running width  & 827.43(23) & 789.29(21) & 748.11(23) & 501.32(21) & 31.715(68) \\
\hline
complex mass   & 826.23(21) & 788.18(21) & 746.78(21) & 500.59(20) & 31.738(68) \\
\hline
\multicolumn{6}{c}{} \\
\hline
\multicolumn{6}{|c|}{$\si(\gamma\gamma\rightarrow\Pem\bar\Pne\Pnmu\Pmup
\ga)$} \\
\hline 
$\sqrt{s_{\ga\ga}}\,[\GeV]$ & 500 & 800 & 1000 & 2000 & 10000 \\
\hline
fixed width   & 39.230(45) & 47.740(73) & 49.781(91) & 43.98(18) & 4.32(23) \\
\hline
step width    & 39.253(45) & 47.781(73) & 49.881(96) & 44.01(18) & 4.31(24) \\
\hline
running width & 39.251(49) & 47.781(74)& 49.898(95) & 44.48(22) & 10.83(28) \\
\hline
complex mass  & 39.221(45) & 47.730(73) & 49.770(91) & 43.97(18) & 4.31(23) \\
\hline
\end{tabular} }
\caption{Cross sections for the processes 
$\ga\ga\rightarrow\Pem\Pnebar\Pnmu\Pmup$ and 
$\ga\ga\rightarrow\Pem\Pnebar\Pnmu\Pmup\ga$ for various CM energies and 
various width schemes without convolution over the photon spectrum.}
\label{tab:width}
\end{table}
The results 
of all four schemes for the process $\ga\ga\rightarrow\Pem\Pnebar\Pnmu\Pmup$ 
agree within the expected accuracy of ${\cal O}(\GW/\MW)$ up to energies 
in the $\TeV$ range. However, for $\ga\ga\rightarrow\Pem\Pnebar\Pnmu\Pmup\ga$ 
the running-width scheme yields totally wrong results for several TeV, while 
the other schemes are still in good agreement.
Although the gauge-invariance-breaking effects in the running-width scheme
are formally of ${\cal O}(\GW/\MW)$, they are enhanced by spoiling
gauge cancellations, thereby ruining the reliability of the prediction
completely.

For the semileptonic $\ggffff$ process it was already observed 
in \citere{Baillargeon:1997se} that the cross section does not
vary significantly if the fixed-width, the running-width, or a so-called
``fudge-factor'' scheme is used for introducing finite widths.

\subsection{Distributions}
\label{se:distr}

\subsubsection{Distributions for reconstructed W~bosons}

\begin{figure}
\setlength{\unitlength}{1cm}
\centerline{
\begin{picture}(7,8)
\put(-1.7,-14.5){\includegraphics{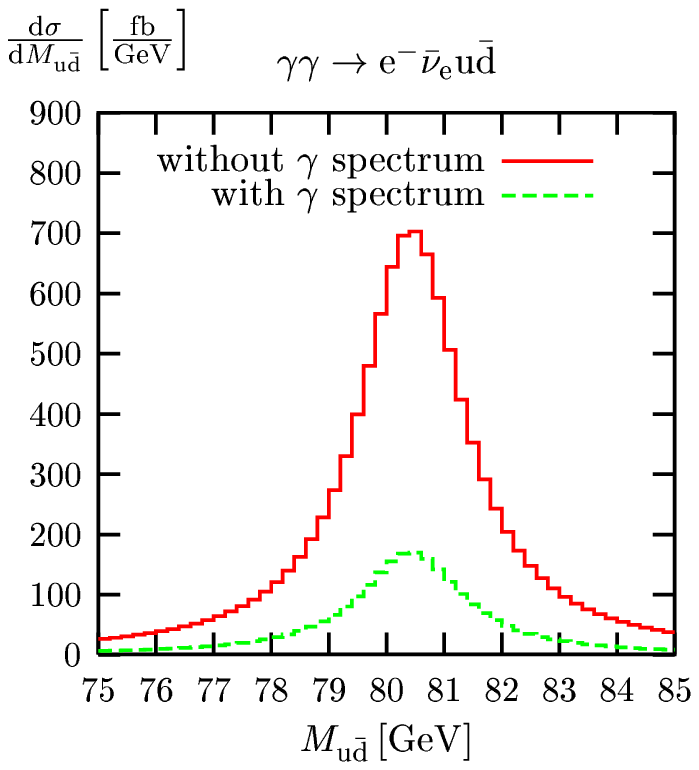}}
\end{picture}\hspace*{1em}
\begin{picture}(7,8)
\put(-1.7,-14.5){\includegraphics{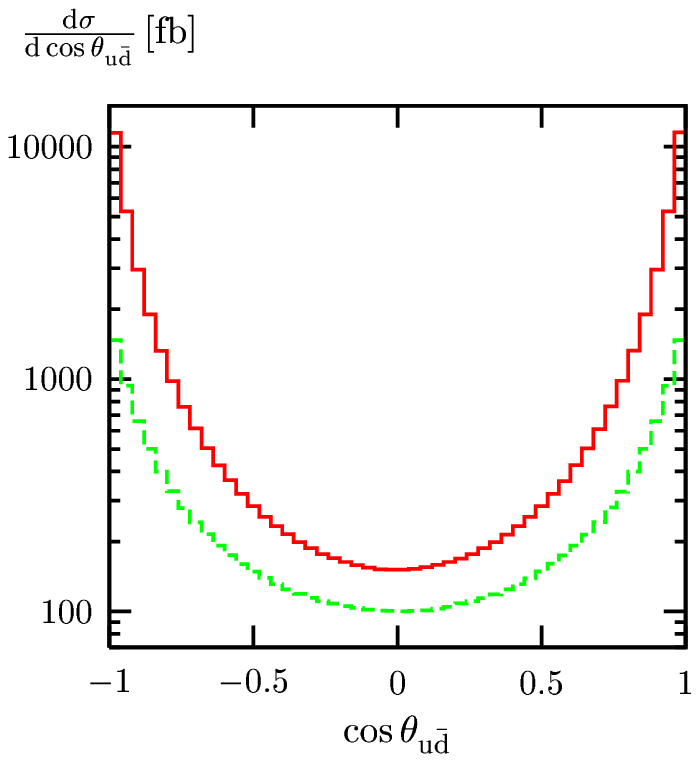}}
\end{picture}}
\caption{Invariant-mass distribution of the $\PW^+$ boson reconstructed 
from the $\Pu\Pdbar$ quark pair (l.h.s.) as well as 
its production-angle distribution 
(r.h.s.) in the reaction $\gamma\gamma\to\Pem\bar\Pne\Pu\bar\Pd$ at 
$\sqrt{s}=500\GeV$
with and without convolution over the photon spectrum.}
\label{fig:wmass}
\end{figure}

In \reffi{fig:wmass} we show the invariant-mass and angular distributions
of the intermediate $\PW^+$ boson for the process 
$\gamma\gamma\to\Pem\bar\Pne\Pu\bar\Pd$ at $\sqrt{s}=500\GeV$. 
The momentum of the $\PW^+$ boson is reconstructed 
from the outgoing quark pair in the decay $\PW^+\to\Pu\bar\Pd$.
\reffi{fig:wmass} also illustrates the effect of the convolution
over the photon spectrum. 

The resonance in the invariant-mass distribution 
(l.h.s.\ of \reffi{fig:wmass}) has the typical 
Breit-Wigner shape and can be used to determine the 
W-boson mass and width at a $\ga\ga$ collider.
Moreover, owing to its large cross section, the W~reconstruction in
this reaction seems to be 
a promising possibility for detector calibration at a $\ga\ga$ collider. 
Similarly to the integrated cross sections 
discussed in the previous sections, the convolution 
qualitatively rescales the distribution by roughly a factor $4$.

The r.h.s.\ of \reffi{fig:wmass} shows the distribution in the angle
$\theta_{\Pu\bar\Pd}$ between the $\PW^+$ boson and the beam axis.
Since the incoming $\ga\ga$ state is symmetric w.r.t.\ interchange of the
two photons, the angular distribution 
is symmetric in the production angle $\theta_{\Pu\bar\Pd}$. 
W~bosons are predominantly produced in 
forward or backward direction owing to diagrams with 
$t$- and $u$-channel exchange of W~bosons.  
For the process $\ggtoww$ with on-shell W bosons, the forward and 
backward peaks are integrable and lead to a constant cross section in the 
high-energy limit.
As already pointed out in \refse{se:energydep}, the angular cuts 
\refeq{eq:cuts} restrict the available phase space of the intermediate 
W bosons and lead to a reduction of the forward and backward peaks 
for sufficiently high energies.
Note that the reduction induced by the convolution over the photon
spectrum is not
uniform, but tends to flatten the shape of the angular distribution
slightly.
This is mainly due to the reduced CM energy in the photon spectrum,
leading to a less pronounced peaking behaviour in the forward and
backward directions.

\subsubsection{Energy and production-angle distributions of fermions}
\label{se:eldistr}

\begin{figure}
\setlength{\unitlength}{1cm}
\centerline{
\begin{picture}(7,7)
\put(-1.2,-13.0){\includegraphics{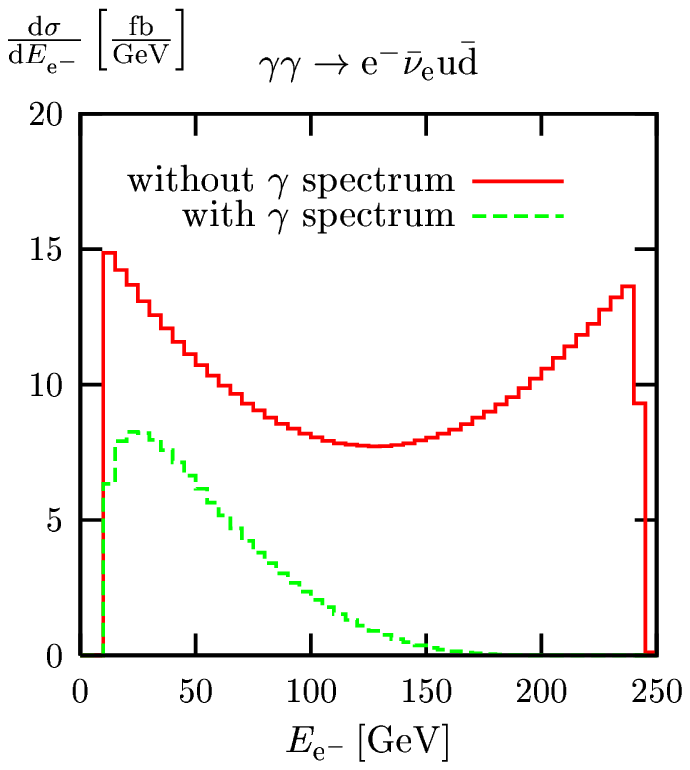}}
\end{picture}\hspace*{1em}
\begin{picture}(7,7)
\put(-1.2,-13.0){\includegraphics{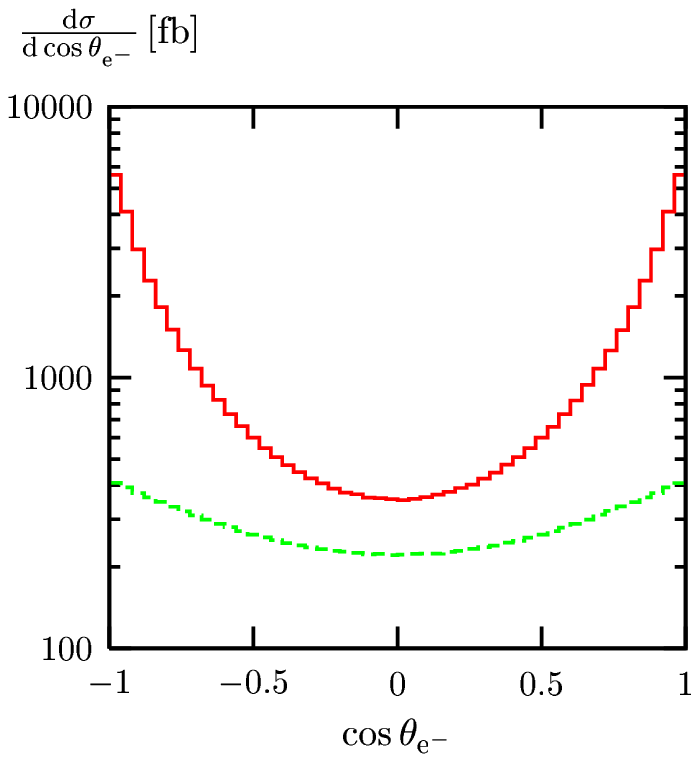}}
\end{picture}
}
\centerline{
\begin{picture}(7,7)
\put(-1.2,-13.0){\includegraphics{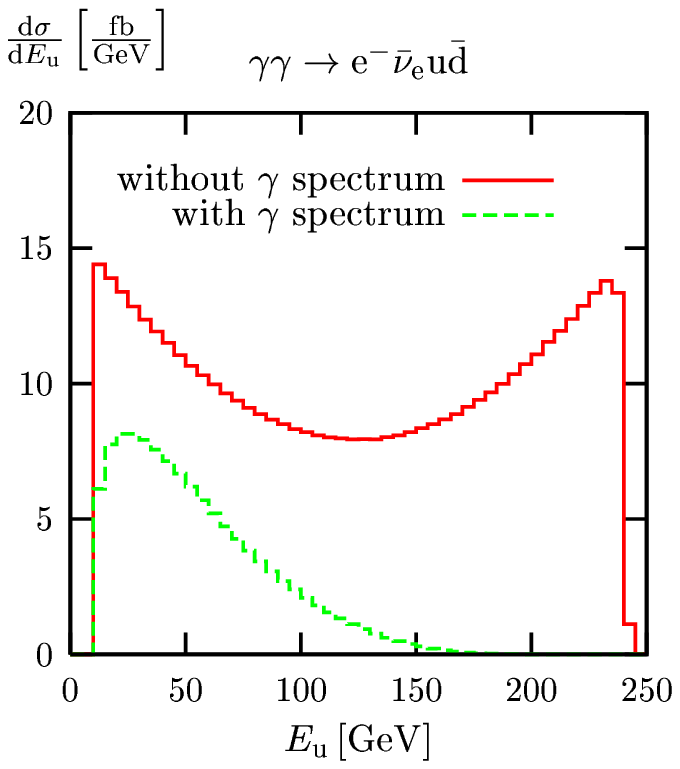}}
\end{picture}\hspace*{1em}
\begin{picture}(7,7)
\put(-1.2,-13.0){\includegraphics{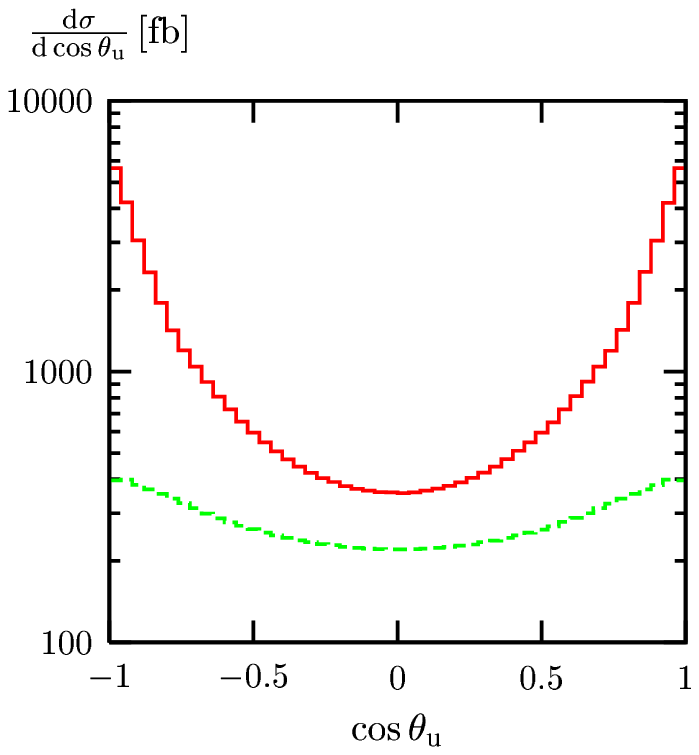}}
\end{picture}
}
\centerline{
\begin{picture}(7,7)
\put(-1.2,-13.0){\includegraphics{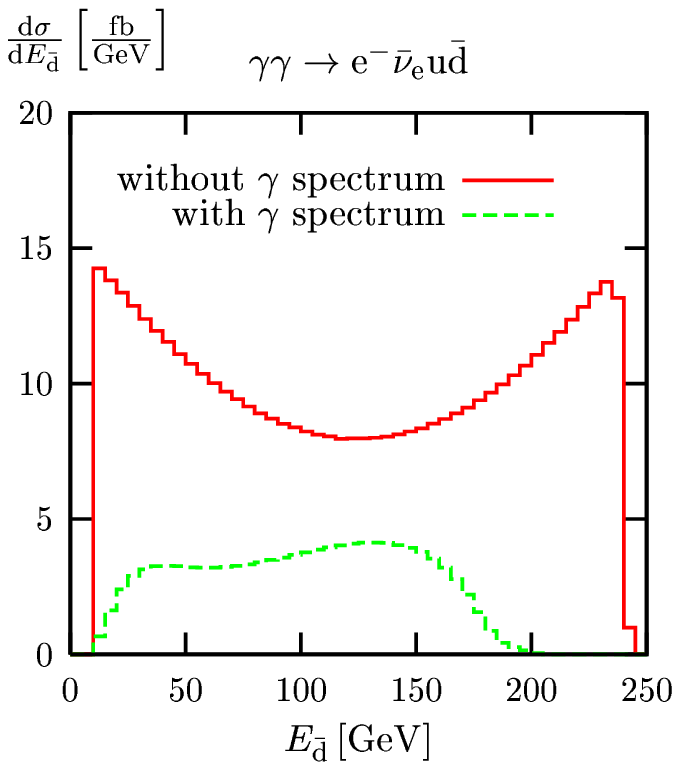}}
\end{picture}\hspace*{1em}
\begin{picture}(7,7)
\put(-1.2,-13.0){\includegraphics{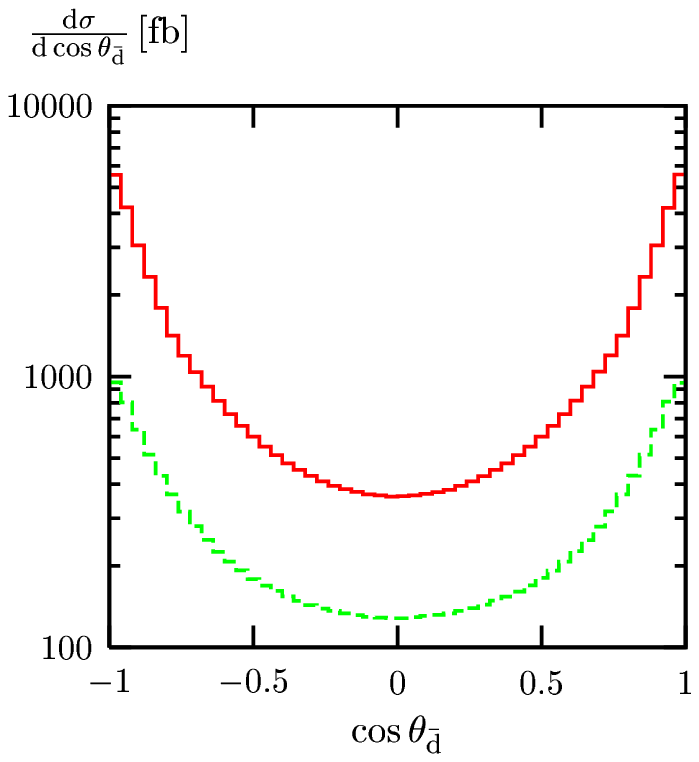}}
\end{picture}
} 
\caption{Energy (l.h.s.) and production-angle (r.h.s.) distributions 
of the outgoing fermions $\Pem$, $\Pu$, and $\bar\Pd$ 
in the process $\gamma\gamma\to\Pem\bar\Pne\Pu\bar\Pd$ at 
$\sqrt{s}=500\GeV$
with and without convolution over the photon spectrum.}
\label{fig:eldistr}
\end{figure}
In \reffi{fig:eldistr} we show the energy and angular distributions
of the outgoing fermions $\Pem$, $\Pu$, and $\bar\Pd$
in the reaction $\gamma\gamma\to\Pem\bar\Pne\Pu\bar\Pd$
at $\sqrt{s}=500\GeV$ with and without convolution over the 
photon spectrum.

For monochromatic, unpolarized incoming $\ga$ beams (i.e.\ without
convolution over the photon spectrum), 
the energy distributions (l.h.s.\ of \reffi{fig:eldistr})
of the fermions $\Pem$, $\Pu$, 
and $\bar\Pd$ almost coincide and are maximal at their 
largest and smallest kinematical limits. These regions are dominated by
the situations where the respective W~boson emits the considered fermion
parallel or anti-parallel to its direction of flight.
The convolution over the photon spectrum changes the shapes of the
energy distributions considerably. 
Since the photon spectrum falls off rapidly for energies above
80\% of the incoming electron energy,
energies of the final-state fermions larger than 
$200\GeV$ become practically impossible.
For fermion energies below $200\GeV$ the shapes of the distributions of
the outgoing fermions $\Pem$ and $\Pu$ look rather different from the
one for the anti-fermion $\bar\Pd$. This effect is due to the
effective $\ga$ beam polarization in the photon spectrum;
for unpolarized $\ga$ beams the energy distributions would look
almost identical.
In detail, the effective polarization of the $\ga\ga$ system is
mainly $(\la_1\la_2)=(++)$, leading predominantly to $\PW^+\PW^-$
production with effective helicities $(++)$
(see, e.g., \citeres{Baillargeon:1997se,Jikia:1996uu}).
Owing to helicity conservation, however, W~bosons with helicity $+1$
cannot decay 
into fermion--anti-fermion pairs
with a fermion (which must have helicity $-\frac{1}{2}$)
parallel to the flight direction of the W~boson. Thus, much more
anti-fermions (which have helicity $+\frac{1}{2}$)
than fermions follow the directions of the decaying W~bosons,
which qualitatively explains the reduction (enhancement) of the 
fermion (anti-fermion)
energy distributions at the upper kinematical energy limit.
The above arguments are nicely illustrated in \citere{Baillargeon:1997se},
where the fermion energy distributions are shown for fully
polarized, monochromatic photon beams.

The r.h.s.\ of \reffi{fig:eldistr} shows the distributions in
the angles $\theta_f$ of the (anti-)fermions $f=\Pem,\Pu,\bar\Pd$
to the beam axis. 
Because of the symmetry of the incoming $\ga\ga$ state 
w.r.t.\ interchange of the two photons, the angular distribution 
is symmetric in $\theta_f$.
The forward and backward peaks originate from two sources.
The by far largest contribution to 
the differential cross section comes from signal diagrams and thus
from configurations where the 
W~bosons as well as the decay fermions 
are nearly parallel to the beam.
The second source,
which is widely suppressed by the applied cuts, 
is related to collinear singularities of background diagrams 
where an incoming photon splits into an fermion--anti-fermion pair
$f\bar f$, with the
fermion or anti-fermion directly going into the final state. 
If the phase space of the outgoing (anti-)fermion is not restricted by cuts,
such collinear or mass singularities lead to logarithms of the form 
$\ln(s/m_f^2)$, where $m_f$ is the fermion mass.
Since our calculation is done for massless fermions, 
the collinear singularities must be excluded by phase-space cuts
and the fermion mass in $\ln(s/m_f^2)$ is replaced by the corresponding 
cut parameter. 

The photon spectrum reduces the differential cross section over the 
whole range and again flattens the angular distributions,
especially in the cases of outgoing fermions.
The significant difference between the outgoing fermions and anti-fermions
is again due to the effective $\ga$ polarization in the photon spectrum.
As explained above, more anti-fermions than fermions follow the flight 
directions of the W~bosons, which are mainly produced in the forward 
and backward directions. This is the reason why the $\theta_{\Pe^-}$
and $\theta_\Pu$ distributions are flattened, while the peaking
behaviour in the $\theta_{\bar\Pd}$ distribution is more pronounced
after the convolution over the photon spectrum.

\subsubsection{Higgs-boson resonance}
\label{se:Higgsres}

\begin{figure}
\setlength{\unitlength}{1cm}
\centerline{
\begin{picture}(7,8)
\put(-1.7,-14.5){\includegraphics{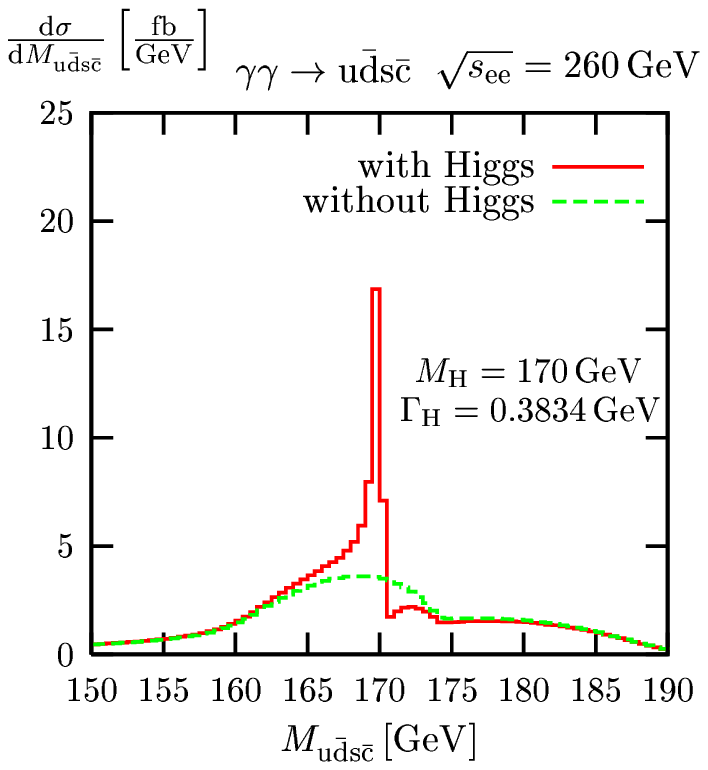}}
\end{picture}
\qquad
\begin{picture}(7,8)
\put(-1.7,-14.5){\includegraphics{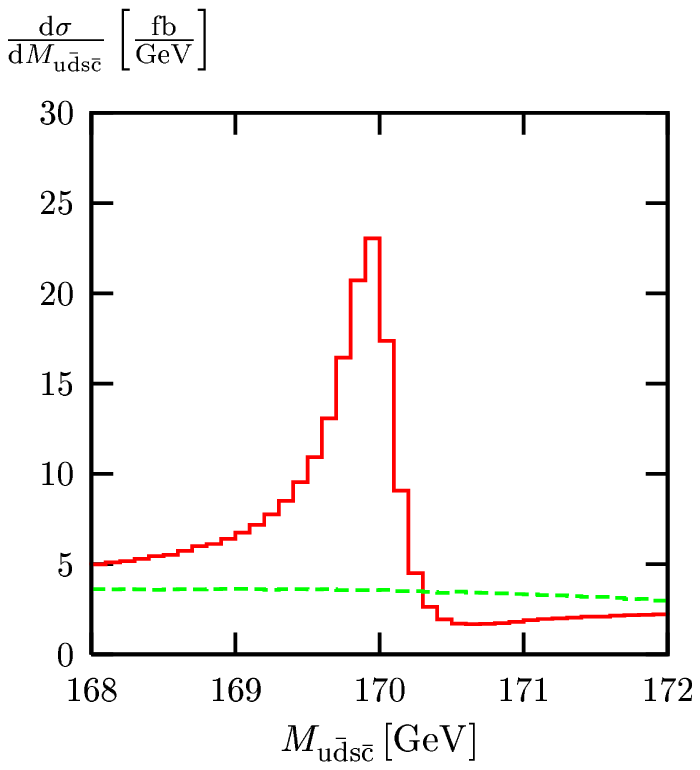}}
\end{picture}} 
\caption{Invariant-mass distribution of the four-quark final state 
for the process $\ga\ga\to\Pu\Pdbar\Ps\Pcbar$ at 
$\sqrt{s_{\mathrm{ee}}}=260\GeV$ including convolution over the
photon spectrum.}
\label{fig:higgsplot}
\end{figure}

In \reffi{fig:higgsplot} we show the invariant-mass distribution of 
the Higgs boson for the process 
$\ga\ga\to\PH\to\PW\PW\to\Pu\Pdbar\Ps\Pcbar$ 
for a Higgs mass of $\MH=170\GeV$.
The CM energy of the electron beams is chosen to be 
$\sqrt{s_{\mathrm{ee}}}=260\GeV$ 
which maximizes the $\ga\ga$ luminosity in the region 
$\sqrt{s_{\ga\ga}}\sim\MH$.
The invariant mass $M_{\Pu\Pdbar\Ps\Pcbar}$
of the Higgs boson is reconstructed from 
its decay products which are the four outgoing quarks.
This means that $M_{\Pu\Pdbar\Ps\Pcbar}$ is equal to the photonic CM
energy, $M_{\Pu\Pdbar\Ps\Pcbar}=\sqrt{s_{\ga\ga}}$.
Thus,
the shape of the distribution depends on the form of the
photon spectrum very strongly.
The effective $\ga\ga\PH$ coupling is set to the SM value \refeq{eq:higgs}.
For comparison the situation without Higgs resonance is also included
in \reffi{fig:higgsplot}, illustrating the significance of the
Higgs signal.
The different peak heights in the two plots simply result from different
bin sizes.

\subsection{Anomalous couplings}
\label{se:AGC}

In this section we study the impact of possible anomalous 
gauge-boson couplings on CC cross sections of the process class 
$\ggffff$. In order to estimate the full
sensitivity of a future $\ga\ga$ collider, such as 
the $\ga\ga$ option at TESLA, on anomalous couplings,
in addition differential distributions and realistic event selections
should be taken into account. Such a study goes beyond
the scope of this paper, but our Monte Carlo generator can serve as
a tool in this task.

\begin{figure}
\setlength{\unitlength}{1cm}
\centerline{
\begin{picture}(7,8)
\put(-1.7,-14.5){\includegraphics{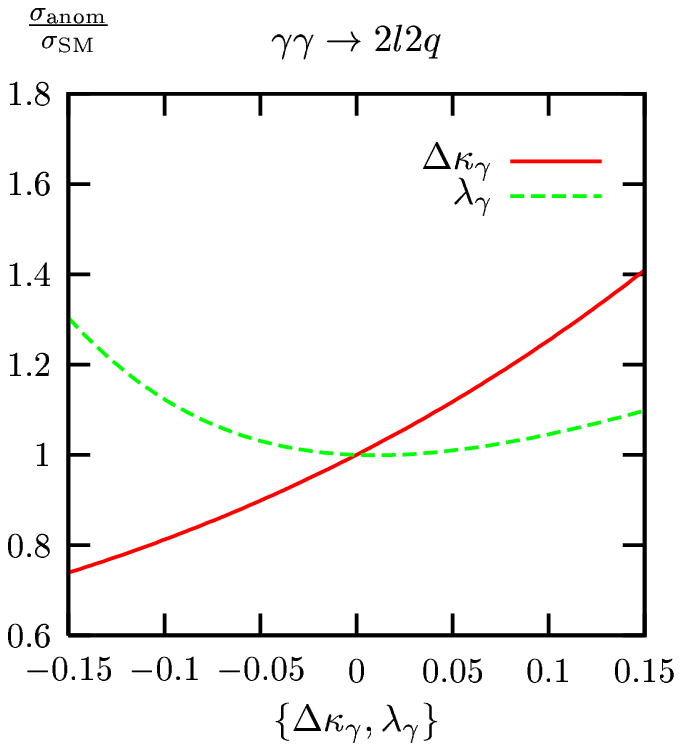}}
\put(.3,-4.3){\includegraphics{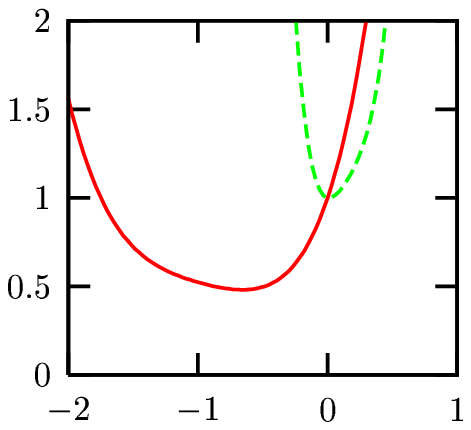}}
\end{picture}\hspace*{1em}
\begin{picture}(7,8)
\put(-1.7,-14.5){\includegraphics{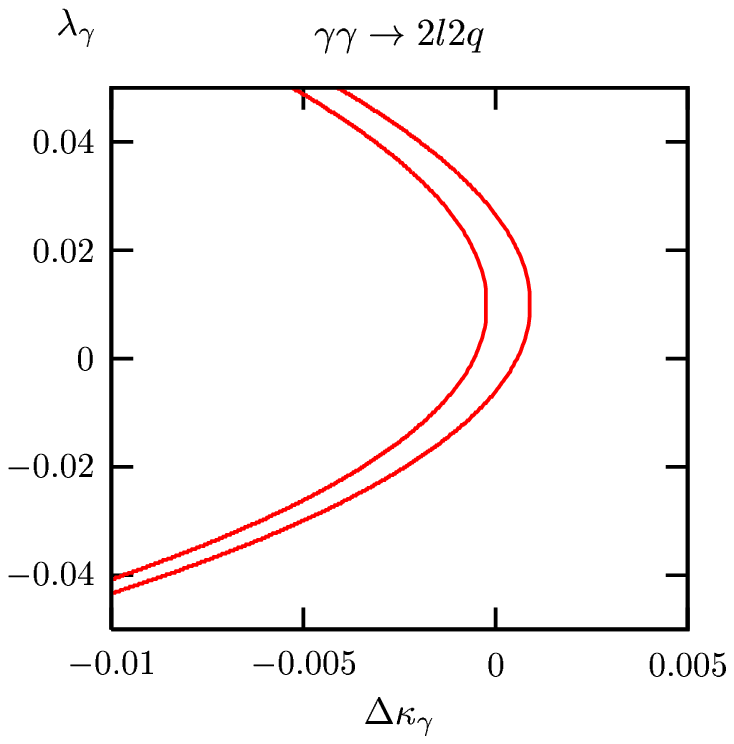}}
\put(.3,-5.2){\includegraphics{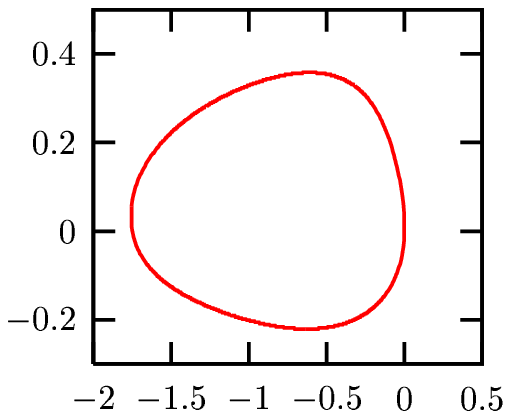}}
\end{picture}}
\caption{Combined cross section for semi-leptonic final states
as a function of the ATGC $\Delta\ka_{\ga}$ and $\la_{\ga}$ (l.h.s.) 
and $1\si$ contours (r.h.s.) in the $(\Delta\ka_{\ga},\la_{\ga})$
plane at $\sqrt{s_{\mathrm{ee}}}=500\GeV$ 
including the convolution over the photon spectrum.}
\label{fig:triple}
\end{figure}

\begin{figure}
\setlength{\unitlength}{1cm}
\centerline{
\begin{picture}(7,8)
\put(-1.7,-14.5){\includegraphics{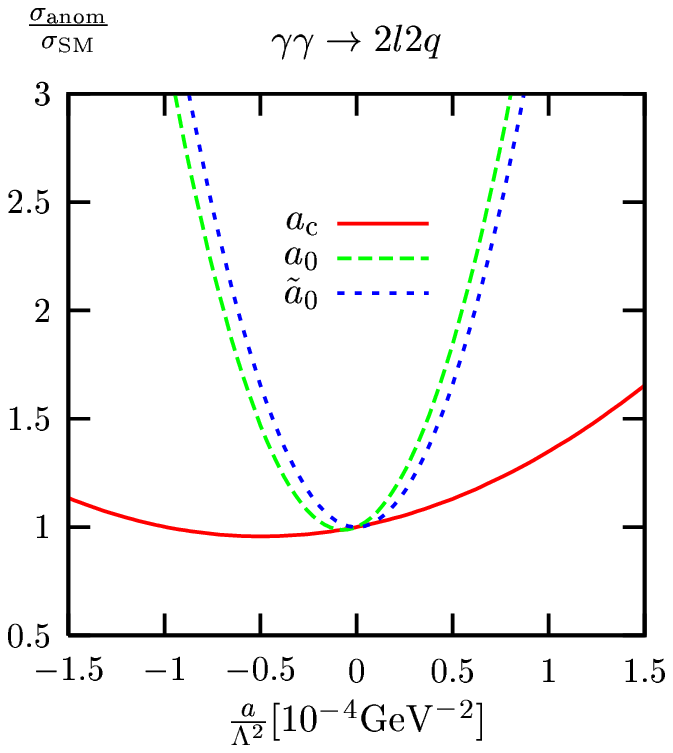}}
\end{picture}}
\centerline{
\begin{picture}(7,8)
\put(-1.2,-14.5){\includegraphics{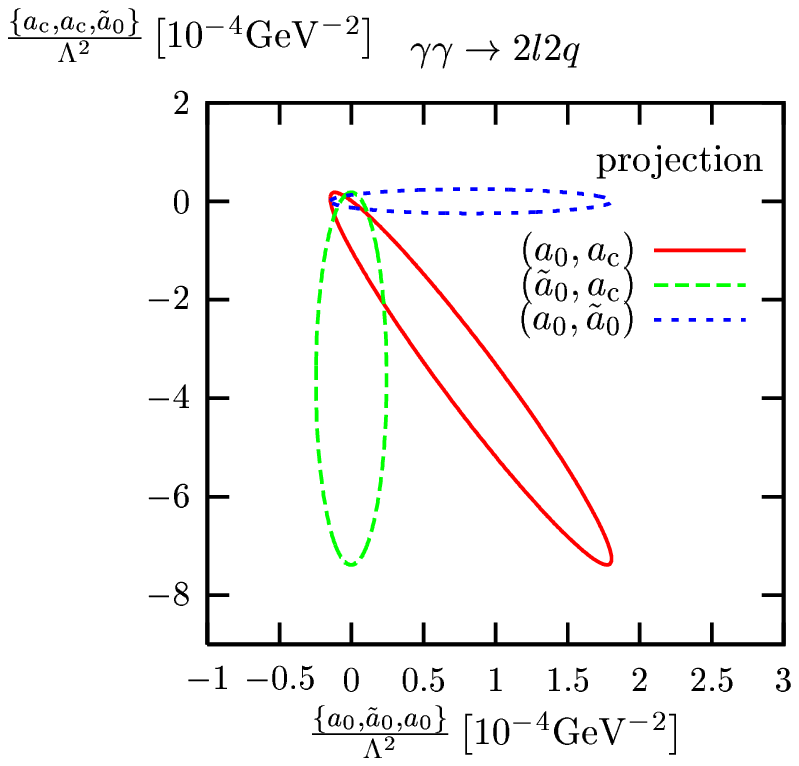}}
\end{picture}\hspace*{1em}
\begin{picture}(7,8)
\put(-1.2,-14.5){\includegraphics{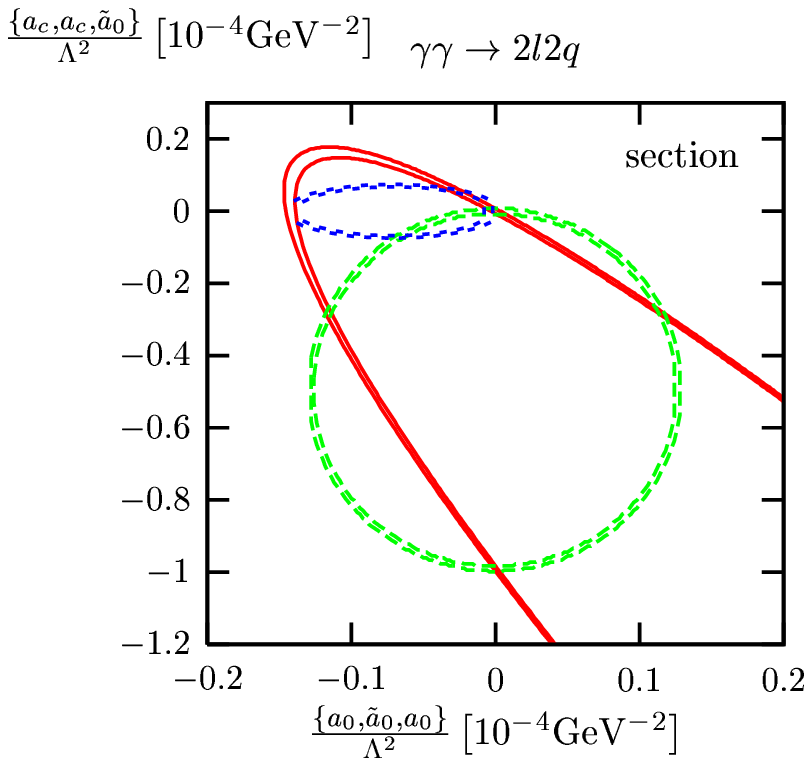}}
\end{picture}}
\caption{Combined cross section for semi-leptonic final states as a 
function of the AQGC $\az$, $\ac$, and $\azt$ (upper plot) and $1\si$ contours 
(l.h.s.\ projection, r.h.s.\ section) in the coordinate planes
at $\sqrt{s_{\mathrm{ee}}}=500\GeV$ 
including the convolution over the photon spectrum.}
\label{fig:quartic}
\end{figure}

We consider only semi-leptonic final states,
since these have the cleanest experimental signal. 
The cross section for semi-leptonic final states is obtained from the sum 
over all reactions $\ga\ga\to l^-\bar\nu_lq\bar q'$, 
with  $q=\Pu,\Pc$ and $l=\Pe,\mu,\tau$, and their corresponding 
charge-conjugated processes $\ga\ga\to \nu_l l^+q'\bar q$. 
The results are shown in \reffi{fig:triple} for ATGC and in 
\reffi{fig:quartic} for AQGC. In the left plot of \reffi{fig:triple} and 
the upper plot of \reffi{fig:quartic} we show the cross section 
as a function of the anomalous coupling constant normalized to the 
SM cross section. As can be seen in the 
insert of \reffi{fig:triple}, the 
minimum in the $\dkag$ curve is shifted to negative values which is caused by 
contributions to the cross section that are linear in $\dkag$. 
These contributions result from the 
interference between matrix elements 
linear in the ATGC $\dkag$ with the SM amplitude.
On the other hand, the interferences for the ATGC $\la_{\ga}$ are small. 
In the case of AQGC, such interferences are relatively large for $\ac$.

In order to examine the sensitivity of a linear collider to anomalous 
couplings, we consider a $\ga\ga$ collider with an integrated
luminosity of $\L=100\fb^{-1}$ 
and a CM energy of $\sqrt{s_{\mathrm{ee}}}=500\GeV$ \cite{Ginzburg:1981vm}.
We define
\beq
\chi^2 \equiv \frac{(N(a_i)-N)^2}{N}
\qquad
\mbox{with} \quad
N=\si_{\mathrm{SM}}\L, \quad
N(a_i)=\si(a_i)\L,
\eeq
where $N$ is the expected number of events in the SM and $N(a_i)$ the number 
of events in the SM extended by the non-standard couplings. 
In \reffis{fig:triple} and \ref{fig:quartic} the $1\si$ contours
corresponding to $\chi^2=1$ are shown. 
Note that the $1\si$ contour can result from $N(a_i)>N$ and $N(a_i)<N$.
While $\chi^2=1$ with $N(a_i)>N$ is always possible for sufficiently
large anomalous couplings, $\chi^2=1$ with $N(a_i)<N$ requires
large interference effects
of matrix elements with anomalous couplings.
In our case, both branches of the $1\si$ contours are realized.
In \reffi{fig:triple} 
the plot on the r.h.s.\ shows the $1\si$ contours in the 
$(\dkag,\lag$) plane. 
Since the cross section is a polynomial 
up to fourth power in the ATGC, the contours are not of elliptic form.
The allowed region lies between the two contours 
that are rather close to each other so that they cannot be 
distinguished in the 
insert which shows the contours on a larger scale. 
Note that in the limit of large luminosity the contour in the insert
of the r.h.s.\ of \reffi{fig:triple} does not shrink to a point,
but reduces to a line in the $(\dkag,\lag)$ plane on which
$\si_{\mathrm{anom}}=\si_{\SM}$. In order to resolve this correlation
between $\dkag$ and $\lag$,
anomalous effects on distributions should be considered, or other
constraints from the $\Pep\Pem$ or $\Pem\ga$ modes 
should be included.

In case of AQGC the cross section is at most quadratic 
in the AQGC, and the $\chi^2=1$ surface 
consists of two ellipsoids in the $(\az,\ac,\azt)$ space. 
The existence of two branches is again due to 
large interferences of anomalous contributions.
In the lower left plot of \reffi{fig:quartic} we show the projections
of the outer ellipsoid into 
the coordinate planes of two AQGC (where the third AQGC is zero).
In the lower right 
plot the sections of both ellipsoids with these planes are given. 
Since the centre of the ellipsoids is shifted in the $\ac$ and $\az$
directions, the terms in $\si(a_i)$ linear in these couplings are 
significant; they result from interferences of the diagram with the
AQGC with the SM amplitude. Interferences that are proportional to
$\azt$ turn out to be small.
From Eq.~\refeq{eq:aqgc} it is obvious that there are no 
$\az\azt$ and $\ac\azt$ terms in $\si(a_i)$. Consequently, the projection
into and the section with the $(\az,\ac)$ plane coincide.
On the other hand, the two other projections and sections differ,
signalling that the $\ac\az$ term in $\si(a_i)$ is significant.

The allowed $1\si$ region ($\chi^2<1$) in the $(\az,\ac,\azt)$ space 
is the shell at the boundary of the shown ellipsoid. 
Similar to the observation made above for the ATGC, 
the size of the ellipsoid does not shrink for larger luminosity,
only the thickness of the shell will decrease. 
This means that the size of the projections shown in the lower left
plot of \reffi{fig:quartic} will not reduce for larger luminosity.
Thus, using only information on an integrated cross section
(for a fixed energy) could not improve the bounds on AQGC w.r.t.\
the ones resulting from $\Pep\Pem\to\PW\PW\ga\to 4f\ga$
\cite{Denner:2001vr}.
However, the thinness of the shell of the ellipsoid, as illustrated
in the lower right plot of \reffi{fig:quartic}, shows that the
bounds can be drastically tightened if the correlation between
the three AQGC is resolved. Differential distributions will
certainly provide this information, 
so that a $\ga\ga$ collider should be able to constrain AQGC
by an order of magnitude better than an $\Pep\Pem$ collider 
operating at comparable energy.

\section{Summary}
\label{se:sum}

A Monte Carlo generator, which is based on multi-channel techniques with 
adaptive self-optimization, 
has been constructed for lowest-order predictions for the processes
$\ggffff$ and $\ggffffg$.
Fermions are treated in the massless approximation.
Anomalous triple and quartic gauge-boson couplings, as well as an
effective $\ga\ga\PH$ coupling, are included 
in the transition matrix elements for $\ggffff$.
For $\ggffff$ compact results for the helicity amplitudes
are presented in terms of Weyl--van-der-Waerden spinor products.

Using this generator, we have presented a variety of
numerical results:
\begin{itemize}
\item
A representative list of integrated 
cross sections for the processes $\ggffff$ and $\ggffffg$ 
is compared to results obtained with the 
{\Whizard}/{\Madgraph} package. 
We find agreement between both Monte Carlo programs. 
\item
The dependence of some $\ggffff$ and $\ggffffg$ cross sections on the CM energy 
is shown. In this context, the influence of a realistic photon beam
spectrum and the size of
subcontributions originating from charged-current,
neutral-current, or strong interactions are investigated.
\item
The complete lowest-order cross section for $\ggwwffff$ is
compared with the corresponding double-pole approximation, 
revealing an accuracy of the latter of $1{-}3\%$. 
While this deviation is of the naively expected order of
$\GW/\MW$, 
the gauge-variant subset of diagrams with 
two resonant W bosons, the so-called ``CC03'' signal diagrams, 
yields differences of $5{-}10\%$ to the full calculation.
The deviations of the naive signal grow with increasing energy,
whereas the quality of the double-pole approximation remains stable.
\item
Different schemes for introducing a finite width 
for gauge bosons are analyzed, and good agreement between the 
gauge-invariant complex-mass scheme and the 
fixed-width scheme is found. 
However, the results also reveal problems (at least in $\ggffffg$)
when using the running width.
\item
A few differential distributions are discussed for $\ggffff$ processes
that proceed via charged-current interactions, 
in particular comprising
distributions in the invariant mass and in the production angle of
reconstructed W~bosons and in the invariant mass of a resonant Higgs
boson.
Moreover, it is shown that the convolution over the photon spectrum
significantly distorts energy and angular distributions of the
produced fermions due to an effective photon polarization.
\item
Finally, we examine the effects of anomalous triple and quartic 
gauge-boson couplings on $\ggffff$ cross sections.
Since contributions of anomalous couplings to cross sections can cancel in
specific configurations, it is necessary to take into account
results from other observables (such as differential distributions)
or from other experiments (such as $\Pep\Pem$ or $\Pem\ga$ collisions)
in order to constrain individual anomalous couplings.
Our results suggest that an analysis of the processes $\ggffff$
can constrain anomalous $\ga\ga\PW\PW$ couplings
about an order of magnitude better than studying $\Pep\Pem\to4f\ga$.
\end{itemize}

Apart from studying the processes $\ggffff$ and $\ggffffg$ in lowest
order the described event generator is a first step towards a
precision calculation of the process $\ggwwffff$ which has to
include radiative corrections to the dominating W-pair production
process. In particular, the results for the radiative processes
$\ggffffg$ can serve as a basic building block in this task.
In a next step, we will complete the calculation of electroweak
corrections in the double-pole approximation
similar to the approach of {\RacoonWW} for $\Pep\Pem$ annihilation.

\section*{Acknowledgement}

A.~Denner is gratefully acknowledged for valuable discussions and
for carefully reading the manuscript.

\end{document}